\documentclass[sigconf]{acmart}

\usepackage{xcolor}  
\usepackage{booktabs}

\usepackage{hyperref}
\usepackage{cleveref}
\crefformat{section}{\S#2#1#3} 
\crefformat{subsection}{\S#2#1#3}
\crefformat{subsubsection}{\S#2#1#3}

\usepackage{lipsum}
\usepackage{pifont}
\newcommand{\cmark}{\ding{51}}%
\newcommand{\xmark}{\ding{55}}%
\usepackage{enumitem}
\usepackage{xspace}
\usepackage{subcaption}
\newcommand{\filler}[1]{\relax}
\definecolor{OwnAzure}{HTML}{336699}
\definecolor{OwnCerulean}{HTML}{CAE2FE}
\definecolor{OwnOliveGreen}{HTML}{556B2F}
\definecolor{KamPurple}{HTML}{907C97}
\usepackage[colorinlistoftodos,prependcaption,textsize=tiny]{todonotes}
\usepackage{xargs}                      
\newcommandx{\todolarge}[2][1=]{\todo[inline,size=\large,linecolor=OwnAzure,backgroundcolor=OwnCerulean,bordercolor=OwnAzure,#1]{#2}}
\newcommandx{\todoai}[2][1=]{\todo[inline,linecolor=OwnAzure,backgroundcolor=OwnCerulean,bordercolor=OwnAzure,#1]{#2}}

\newcommandx{\addref}[2][1=]
{\todo[inline,linecolor=blue,backgroundcolor=blue!50,bordercolor=blue,#1]{Add reference. #2}}
\newcommandx{\unsure}[2][1=]{\todo[inline, linecolor=red,backgroundcolor=red!25,bordercolor=red,#1]{#2}}
\newcommandx{\change}[2][1=]{\todo[inline, linecolor=blue,backgroundcolor=blue!25,bordercolor=blue,#1]{#2}}
\newcommandx{\improvement}[2][1=]{\todo[linecolor=Plum,backgroundcolor=Plum!25,bordercolor=Plum,#1]{#2}}
\newcommandx{\thiswillnotshow}[2][1=]{\todo[disable,#1]{#2}}

\definecolor{darkred}{rgb}{0.5,0,0}
\definecolor{darkgreen}{rgb}{0,0.5,0}
\definecolor{darkblue}{rgb}{0,0,0.5}

\graphicspath{{images/}{figures/}}

\usepackage{glossaries}
\newacronym{mmog}{MMOG}{massively multiplayer online game}

\newacronym{mve}{MLG}{Minecraft-like Game}
\newcommand{\mve}{\gls{mve}\xspace}
\newcommand{\mvepl}{\glspl{mve}\xspace}
\newcommand{\mves}{\mvepl}

\newcommand{\das}{DAS-5\xspace}

\newacronym{qos}{QoS}{Quality of Service}
\newcommand{\qos}{\gls{qos}\xspace}

\newcommand{\envworkload}{environment-based workload\xspace}
\newcommand{\envworkloads}{environment-based workloads\xspace}
\newcommand{\Envworkloads}{Environment-based workloads\xspace}

\newcommand{\toolname}{Meterstick\xspace}

\newacronym{minecraftlike}{MVE}{modifiable virtual environment}
\newacronym{dc}{dyconit}{dynamic consistency unit}
\newacronym{aoi}{AoI}{area of interest}
\newacronym{is}{IS}{interest set}
\newglossaryentry{solution}{
  name={Dyconit},
  description={}
}
\newacronym{npc}{NPC}{Non-Playable Character}
\newcommand{\npc}{\gls{npc}\xspace}
\newcommand{\npcs}{\glspl{npc}\xspace}

\newacronym{fps}{FPS}{frames per second}
\newcommand{\fps}{\gls{fps}\xspace}

\newacronym{ISR}{ISR}{Instability Ratio}
\newcommand{\newmetric}{\gls{ISR}\xspace}
\newcommand{\newmetricfull}{Instability Ratio\xspace}

\DeclareRobustCommand{\designref}[1]{%
  \begin{tikzpicture}[baseline=(char.base)]
    \node[draw,circle,inner sep=0.5pt, fill=black, text=white] (char){\small #1};
  \end{tikzpicture}%
}

\newcommand{\vcutS}{\vspace*{-0.15cm}}
\newcommand{\vcutM}{\vspace*{-0.25cm}}
\newcommand{\vcutL}{\vspace*{-0.5cm}}

\newenvironment{myitemize}[1][]
{
  \setlength{\leftmargini}{10pt} 
  \setlength{\parskip}{0pt} 
  \begin{enumerate}[#1]
    \setlength{\itemindent}{0pt} 

          \setlength{\itemsep}{0pt} 
          \setlength{\parsep}{0pt} 
          \setlength{\parskip}{0pt} 
          }
          { \end{enumerate}
}

\input{setup-comments}

\setreviewsoff

\copyrightyear{2023}
\acmYear{2023}
\setcopyright{acmlicensed}
\acmConference[ICPE '23] {Proceedings of the 2023 ACM/SPEC International Conference on Performance Engineering}{April 15--19, 2023}{Coimbra, Portugal.}
\acmBooktitle{Proceedings of the 2023 ACM/SPEC International Conference on Performance Engineering (ICPE '23), April 15--19, 2023, Coimbra, Portugal}
\acmPrice{15.00}
\acmISBN{979-8-4007-0068-2/23/04}
\acmDOI{10.1145/XXXXXX.XXXXXX}

\begin{document}

\title{Meterstick: Benchmarking Performance Variability in Cloud and Self-hosted Minecraft-like Games Technical Report}

\author{Jerrit Eickhoff}
\email{J.D.Eickhoff@student.tudelft.nl}
\orcid{0000-0003-0627-294X}
\affiliation{%
  \institution{Delft University of Technology}
  \city{Delft}
  \country{Netherlands}
}

\author{Jesse Donkervliet}
\email{J.J.R.Donkervliet@vu.nl}
\orcid{0000-0002-3067-6402}
\affiliation{%
  \institution{Vrije Universiteit Amsterdam}
  \city{Amsterdam}
  \country{Netherlands}
}

\author{Alexandru Iosup}
\email{A.Iosup@vu.nl}
\orcid{0000-0001-8030-9398}
\affiliation{%
  \institution{Vrije Universiteit Amsterdam}
  \city{Amsterdam}
  \country{Netherlands}
}


\begin{abstract}
  Due to increasing popularity and strict performance requirements, online games have become a
workload of interest for the performance engineering community.
One of the most popular types of online games is the \mve, in which players can terraform the environment. The most popular \mve, Minecraft, provides not only entertainment, but also educational support and social interaction, to over 130\,million people world-wide.
\mves currently support their many players by replicating isolated instances that support each only up to a few hundred players under favorable conditions.
In practice, as we show here, the real upper limit of supported players can be much lower.
In this work, we posit that performance variability is a key cause for the lack of scalability in \mves.
We propose a novel operational model for \mves
and use it to design the first benchmark that focuses on \mve performance variability, defining specialized workloads, metrics, and processes.
We conduct real-world benchmarking of \mves
and find \envworkloads and cloud deployment to be significant sources of performance variability: peak-latency degrades sharply to 20.7~times the arithmetic mean, and exceeds by a factor of 7.4~the performance requirements.
We derive actionable insights for game-developers, game-operators, and other stakeholders to tame performance variability.

\end{abstract}

%





\settopmatter{printacmref=false, printccs=false}
\renewcommand\footnotetextcopyrightpermission[1]{}
\pagestyle{plain} 
\maketitle

\section{Introduction}
\label{sec:intro}

\begin{figure}[t]
    \centering
    \includegraphics[width=\linewidth]{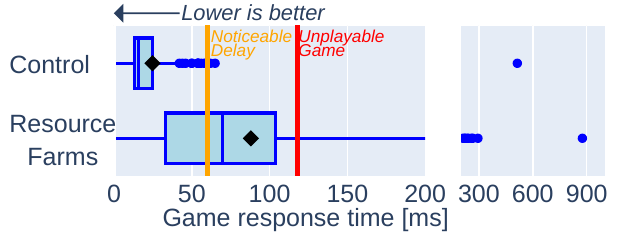}
    \vspace*{-0.65cm}
    \caption{Minecraft response time in the AWS cloud.}
    \label{fig:preview_result}
    \vspace*{-0.5cm}
\end{figure}

The gaming industry is the world's largest entertainment industry~\cite{gamingbiggestindustry}---world-wide, games engage over 3~billion players and yield over \$170~billion in revenue~\cite{newzoo2021}.
In this work, we focus on \emph{Minecraft-like Games (\mves)}, an emergent and highly popular type of game where users can change almost every part of the environment.
The canonical example of an \mve is Minecraft, which is already the best-selling game of all time~\cite{Madan2019May}.
All \mves, including Minecraft, present an important challenge to the performance engineering community: although their user-bases can exceed 100 million active users per month\remove{\footnote{Minecraft has over 130\,million active users/month~\cite{2windows2020}, more than, e.g., MacOS~\cite{Warren2017Apr}.}}, their \textit{scalability} is limited to only 200-300 players even under very favorable conditions~\cite{yardstick}. (\mves support high concurrency by creating separate replicas of their virtual worlds, essentially sharding state and not allowing cross-instance interaction.)
\textit{What limits \mve scalability?} 
In this work,
we posit performance variability is a key limit to \mve scalability,
and design and use an \mve benchmark focusing on this concept.

\mves represent an important and unique class of online multiplayer games.
Most importantly, \mves allow players to create, modify, and remove in-game objects (e.g., player apparel) and geographical features (e.g., terrain)~\cite{DBLP:conf/hotcloud/DonkervlietTI20}.
Moreover, some game objects and features are self-acting, that is, they act even when no player input is applied to them. Players can use them to create dynamic elements, by ``programming'' the environment with combinations of self-acting parts.

\textit{Performance variability} prevents \mve service providers from giving strict \qos guarantees,
and simultaneously incentivizes overprovisioning of resources and limiting
the number of players that can interact together.
For example, Minecraft Realms, a Minecraft \textit{cloud-based service} offered by Microsoft, limits the 
number of players per game-instance to at most 10~(ten)\textbf{!}
In contrast, Hypixel, at 216,762 online players~\cite{minetrack} the most populated Minecraft server, achieves high player-count by stitching together thousands of (independent) \mve instances using specialized tools, but players in different instances cannot interact.

In this work, \emph{we show for the first time empirical evidence that current \mves experience significant performance variability}.
Figure~\ref{fig:preview_result} depicts an exemplary result---
even with a single connected player, the response time varies from good~(below 60\,ms) to \textit{unplayable}~(above 118\,ms). We discuss this and similar real-world experiments in \Cref{sec:mf1}.

Furthermore, \textit{ours is the first study to systematically analyze the effects of performance variability on the operation of \mves.} By designing and using for this purpose a novel benchmark, we provide an important complement to an emerging body of knowledge.
Game researchers and engineers are already aware of the impact of several types of performance variability.
Performance variability in networks causes players to stop playing sooner~\cite{DBLP:journals/cacm/ChenHL06},
and there are widespread techniques in industry to prevent variability in frame rates~\cite{UnityMan69:online,DynamicR99:online}.
However, the effect of performance variability on the interactive simulation of virtual worlds, and in particular on \mves, is much less understood.

Prior work in understanding the performance of \mves~\cite{yardstick,DBLP:journals/corr/abs-1907-13440} and on improving their scalability~\cite{DBLP:conf/netgames/DiaconuKV13,DBLP:conf/netgames/EngelbrechtS13,DBLP:conf/icdcs/DonkervlietCI21} forms a valuable contribution to the field, but does not currently consider explicitly performance variability.
Addressing this important gap, we make a four-fold contribution:

\begin{myitemize}[label=\textbf{C\arabic*}]
    \item\label{contrib:model} We propose an operational model of \mves.
    Ours is the first to consider \mve-specific workloads (\Cref{sec:mve-workload-model}).
    Because \mves allow players to program the virtual environment and terraform the terrain, they support new types of workload not available in most traditional online games.

    \item\label{contrib:design} We design \toolname, a benchmark that quantifies performance variability in \mves (\Cref{sec:design}).
    To this end, we propose a novel performance variability metric, and define a benchmarking approach to produce it experimentally.
    Our benchmark
    supports common deployment-environments for \mves offered as a service, in particular, both cloud-based and self-hosted.
    Our benchmark is the first to quantify performance variability in \mves.

    \item\label{contrib:experiments} We conduct real-world experiments using \toolname~(\Cref{sec:experiments}) and, after analyzing the results, propose actionable insights~(\Cref{sec:actionable_insights}). We evaluate the performance variability of three popular \mves, running on two popular commercial cloud providers and one local compute-cluster.


    \item\label{contrib:FAIR} Following open-science and reproducibility principles, we publish {Findable, Accessible, Interoperable, and Reusable (FAIR~\cite{data:FAIR16})} data, available on Zenodo~\cite{meterstick:data:orig}, and Free-access Open-Source Software (FOSS) artifacts, available on GitHub~\cite{meterstick:git}. 
\end{myitemize}

\section{Operational Model of Minecraft-like Games}
\label{sec:mve-workload-model}
\label{sec:model}

For contribution~\ref{contrib:model}, first, we summarize a state-of-the-art 
operational model and a reference architecture for \mves~(\Cref{sec:model:mve-archi}).
Second, we define \mve-specific \emph{\envworkloads} that are caused by terrain and entity simulation; \Cref{sec:model:workloads} defines the resulting \mve workload model.
Third, 
we model the operational elements of these workloads~(\Cref{sec:model:operational}).


\subsection{Reference Architecture for MLGs}
\label{sec:model:mve-archi}

\begin{figure}[t]
    \includegraphics[width=1\linewidth]{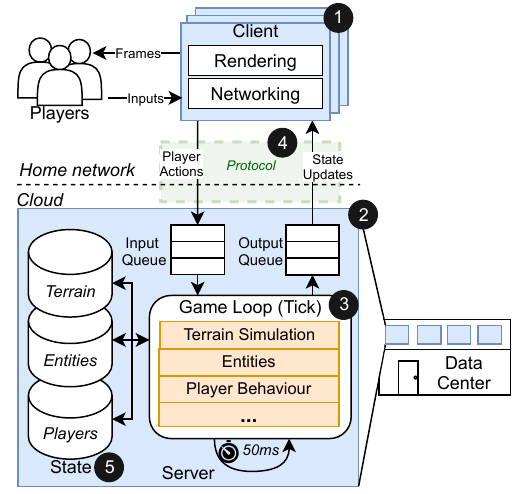}
    \vcutL{}\vcutS{}
    \caption{Overview of an \mve.} 
    \label{fig:mve}
    \vcutL{}
\end{figure}

We leverage in this work a common reference architecture for \mves~\cite{yardstick}.
%
%
As Figure~\ref{fig:mve} depicts,
\mves use a client-server architecture and are commonly deployed in cloud environments.
Players run a client on their own device, which connects to a server instance running in the cloud.
Some \mve developers publicly distribute their server software, allowing players to self-host game instances.
Popular cloud providers such as Amazon~AWS and Microsoft~Azure provide tutorials for running these servers on their platform.
Microsoft, the company that currently owns Minecraft, markets it as a cloud-based service through \emph{Minecraft Realms}, which offers players a fully-managed Minecraft instance for a monthly fee~\cite{Realms}.
Additionally, many smaller companies offer \mves as a service; an extensive 
list appears in 
\Cref{sec:experimental_setup}.

The client~(\designref{1}) has two main tasks. First, it translates player input into in-game actions, which it speculatively applies to the local state and also sends to the server for validation.
The client-server communication uses an implementation-specific protocol~(\designref{4}) that can be shared between different \mves.
Second, the client visualizes the game state, at a fixed rate.


The server~(\designref{2}) is responsible for performing all in-game~(virtual-world) simulations, maintaining the global
state, and disseminating state-updates to clients.
Different from simulators in science or engineering, video game simulations tolerate (temporary) inconsistency, and must support modifying the environment via user input.
The \emph{game loop}~(\designref{3}) performs simulations, by applying state-updates to the global state in discrete steps (\emph{ticks}), at a fixed frequency.
In \mves, this frequency is typically set to 20\,Hz, or 50\,ms per tick.
If a tick takes under 50\,ms, the \mve waits for the next scheduled tick start. However, if a tick exceeds 50\,ms, the tick frequency drops below 20\,Hz and the server enters an \emph{overloaded} state.
While in this state, \textit{the game fails to meet its \qos requirements} and can cause players to experience game stuttering, visual inconsistency, 
and increased input latency.
Prior work has shown direct causality between increased input latency and reduced player experience~\cite{claypoollatency, unplayable118ms, latency60ms}.

\mves generate workloads, both data- and compute-intensive, that do not exist in other types of games.
In contrast to traditional games, \mves allow modifications to the terrain.
This requires the game server to simulate terrain changes and manage terrain-state alongside the player- and entity-state found in traditional games~(\designref{5}).
Unlike
other types of state, 
terrain state can be both \textit{data-intensive} and \textit{compute-intensive}, without direct player input.

\subsection{Workloads in MLGs}\label{sec:Workloads}\label{sec:model:workloads}
\begin{figure}[t]
    \includegraphics[width=\linewidth]{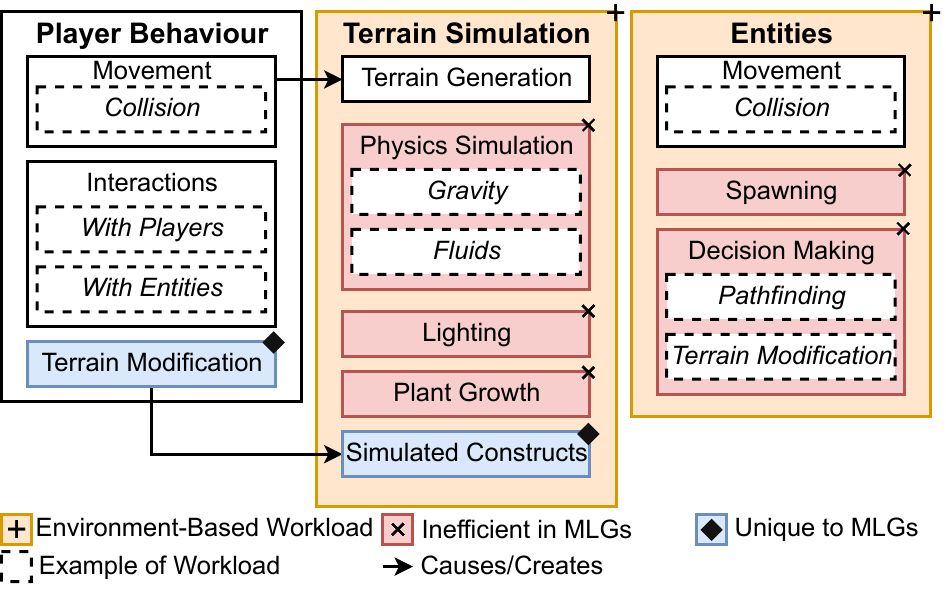}
    \caption{Workload components in \mves.}
    \label{fig:workloads}
    \vcutM{}
\end{figure}

This section presents our workload model for \mves, which focuses on players, terrain, and entities.
We discuss each of these components, in turn, focusing on unique and challenging aspects.
We distinguish 
novel aspects in our research.

Figure~\ref{fig:workloads} presents a visual overview of our model.
Beyond the state-of-the-art, our workload model captures \emph{\envworkloads}, which are caused by simulating the modifiable environment itself, and scale independently from the number of active players. We argue \envworkloads are an important part of representative benchmarking workloads for \mves. However, existing benchmarks do not include this type of workload. Addressing this gap, we propose an \mve workload model which describes a wide range of \envworkloads. In Figure~\ref{fig:workloads}, Terrain Simulation and Entities are examples of \envworkloads. 


\subsubsection{Workload from Players~(known)}
\label{sec:model:workload:players}

Players cause workload for \mves, and games in general, through their actions.
\mves support player-actions found in traditional games, e.g., player movement and interactions, and also \textit{\mve-specific actions}, e.g., to modify terrain.
For player movement, the game computes collisions to prevent players from walking through obstacles such as walls, and disseminates location-changes to other players.
Players can also interact with other players and entities (i.e., objects), for example by collecting resources and exchanging them with other players.

An important difference between \mves and traditional games is support for player-actions that modify the terrain.
In \mves, players can \textit{terraform}---create, modify, and destroy the terrain, as well as the buildings standing on the terrain.
This can generate resource-intensive workloads in two ways.
First, players can change a large part of the terrain in a short amount of time, for example through the use of explosives.
This is both compute- and data-intensive, because the game needs to compute the new terrain, and communicate state updates to keep a consistent view across all players.
Second, players can construct dynamic elements such as \emph{simulated constructs}, which increase the complexity of the terrain simulation
and are discussed in~\Cref{sec:model:workloads:environment}.
The impact of player workloads has been previously examined in both the context of traditional video games architectures and \mves specifically~\cite{MMOPlayersWorkload,yardstick}.

\begin{figure*}[t]
    \includegraphics[width=\textwidth]{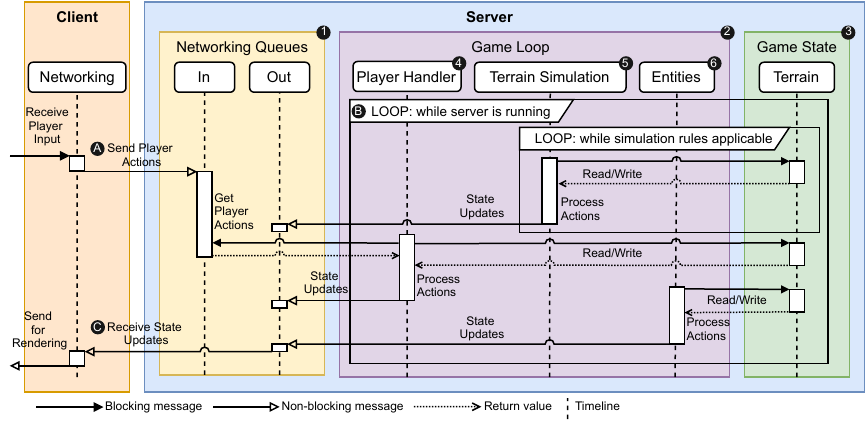}
    \vcutL{}
    \caption{Operational model of an \mve. Note different elements of the Game Loop can be run concurrently.}
    \label{fig:operational}
    \vcutM{}
\end{figure*}

\subsubsection{Workload from Terrain Simulation~(novel)}
\label{sec:model:workloads:environment}

In contrast to traditional games, a significant part of the \mve workload can come from generating and simulating the terrain.
\mves typically present players with an endless open world.
This \textit{world} is split into areas, which are lazily generated when players come near them.
Once the terrain is generated, the game simulates it and allows players to modify it.

We identify four important components of terrain simulation: physics, lighting, plant growth, and simulated constructs.
Although \textit{physics and lighting} simulations are present in traditional games,
the modifiable nature of the terrain makes it significantly more challenging to manage such features in \mves.
Unlike static environments, where physics simulation only needs to happen for the relatively few entities that can move through the world, \mves need to perform physics simulations on the many blocks that compose the terrain itself.
For example, a bridge can collapse when a player removes its support pillars, or the terrain underneath them.
Once the bridge has collapsed, the bridge no longer casts shadow, so the simulator needs to recompute lighting (frequently) at runtime; static environments do not have this dynamic workload.

\textit{Plant growth} is an example of a dynamic element unique to \mves. 
Plants and trees change over time, reshaping the nearby terrain, thus generating new workload.

Through terrain modification, players can create \emph{simulated constructs}.
In a simulated construct, players place together dynamic elements (e.g., plants, automatic croppers) to achieve a certain goal.
For example, many players build irrigation systems that grow and harvest vegetables automatically, with high yield.
Such systems can leverage tens to hundreds of dynamic elements,
whose interaction generates compute-intensive workload for terrain simulation.


\textit{In \mves}, as we show in~\Cref{sec:experiments}, \textit{even a single player can overload the game simulator.}
This is in part because, in \mves, a single player can trigger complex simulations, for example, by building simulated constructs of arbitrary size.
By contrast, in traditional games, only the number of concurrent players is strongly correlated with workload intensity.



\subsubsection{Workload from Entities~(novel)}

An \emph{entity} is an object that exists in the virtual world but is not a player or terrain.
Examples include \npcs, mobiles (i.e., \emph{mobs}), and items (e.g., a sword).
Entities can typically move or be moved by players and collide with each other.
Here we describe two important aspects of entity simulation which are challenging for \mves.

First, games typically instantiate entities at \emph{spawn points}, e.g.,
spawn an \npc at a spawn point in a dark cave when a player is about to enter.
In contrast to static environments, where game developers typically place these spawn points manually, \mves need to compute spawn points dynamically
as
terrain modification may obstruct existing spawn points.

Second, \npcs use path-finding algorithms to move around the map.
Static worlds pre-compute overlay graphs with viable \npc locations,  improving computational efficiency.
In contrast, \mves have changing terrain, so they must compute path-finding graphs dynamically, leading to additional compute-intensive workload.
\subsection{Operational Model of \mves}
\label{sec:model:operational}

We detail in this section the game loop used by \mves.
We
define the \emph{operational model} as the set of operations, and of events triggering and linking them, of individual components in the  implementation of the game loop.
Novel, in this work, we analyze the performance implications of the unique aspects of \mve workloads~(see~\Cref{sec:model:workloads}) when executed with the \mve operational model.

Figure~\ref{fig:operational} depicts a constructed, generalized, and simplified operational model of the \mve game loop. 
 To run the game loop, the game server orchestrates primarily three main components, the
Networking Queues~(component~\designref{1} in Figure~\ref{fig:operational}), the Game Loop~(\designref{2}), and the Game State~(\designref{3}), which we discuss in turn.




The Networking Queues~(\designref{1}) buffer between the game clients and the server.
When a client sends a player-action to the server, it is buffered in the incoming network queue until the next tick.
When the server needs to send a state-update to one or several clients,
it forwards the message to the networking queues, to be further buffered in the outgoing queue or sent directly to the client.

The Game Loop~(\designref{2}) simulates the virtual world and is the core of the game server.
In an \mve, the game loop consists of three elements: players, the terrain, and entities. These elements correspond to the workloads specified in~\Cref{sec:model:workloads}.
Figure~\ref{fig:operational} shows each of these elements, and how they differ from the others.
For each element, its simulation typically requires reading the current game state~(\designref{3}), and may result in terrain state changes that need to be persisted~(i.e., written).
In this section, we focus on the terrain state because it is idiosyncratic to \mves.
Below, we discuss each simulation element in turn.


The Player Handler~(\designref{4}) is driven by player actions, which the Game Loop retrieves from the Networking Queues once per tick.
For example, players can move or build something in the virtual world.
Because the terrain can obstruct the player from performing these actions, the Player Handler must read the terrain state in the vicinity of the player.
If the action is successful, player actions that affect the terrain (e.g., building) need to be written back to the global Game State.

Terrain Simulation~(\designref{5}) is largely independent from player input, and is instead driven by terrain state updates. When a terrain state update occurs, the Terrain Simulation applies its simulation rules to the new state. For example, a terrain simulation rule such as \emph{if terrain is not physically supported, it falls down} can be triggered when a player removes the keystone from a bridge. These rules trigger in a loop, where each iteration informs the adjacent terrain that it is no longer supported. The resulting state changes are written back to the global Game State.

Entities~(\designref{6}) are primarily driven by the Game State, including the state of the terrain, players, and entities themselves.
Entities such as \npcs need the terrain state primarily for pathfinding. In some cases, entities may themselves modify the terrain. For example, an \npc may place or remove terrain, or items such as explosives may remove large parts of terrain all at once.

Although, from a performance perspective, it is desirable to run the game loop elements concurrently,
there are two challenges with this approach.
First, while these elements are in principle independent, they have implicit dependencies through the game state which they access.
Individual elements can only run concurrently as long as they do not access the same state.
Second, terrain simulation rules can cause a sequence of state changes which cannot be parallelized, as is the case in the bridge example above.

Using our operational model for \mves, we formulate two implications for \mve performance variability.
First, because \envworkloads do not rely on the presence of players, large \envworkloads can cause ticks to exceed their maximum duration,
even with few or no players connected.
Second, because player simulation and \envworkloads must be completed sequentially when they access the same state,
even small \envworkloads can affect tick duration given they are spatially clustered.

\section{\toolname Benchmark Design}
\label{sec:design}\label{sec:system-design}

To address contribution~\ref{contrib:design}, we design \toolname, a benchmark for evaluating performance variability in \mves. The main novelty of \toolname relates to its workloads (\Cref{sec:benchmark_workloads}) and performance metrics (\Cref{sec:design:metrics} and~\Cref{sec:metric}).

\subsection{System Requirements} \label{sec:requirements}

Here we describe our eight requirements for \toolname.
We define the first three requirements specifically for our use case.
The last five relate to benchmarking computer systems in general, and are based on existing guidelines~\cite{DBLP:conf/performance/Weicker02, performanceanalysis}.
\begin{enumerate}[label=\textbf{R\arabic*}]
  \item\label{req:capture-variability} \textit{Captures performance variability of \mves:} \toolname must be capable of capturing relevant performance metrics at a granularity sufficient for analysis of variability. The specific measure of variability must be applicable and meaningful in the context of \mves.

  \item\label{req:valid-workloads} \textit{Validity of workloads:} The workloads used in benchmarking of the \mve should be representative of real-world use and address the workload types listed in~\Cref{sec:Workloads}.

  \item\label{req:extensive-exps-and-metrics} \textit{Relevant metrics and experiments:} The benchmark should support relevant experiments to isolate different sources of variability, and collect meaningful metrics to allow suitable analysis of these sources.

  \item\label{req:fairness} \textit{Fairness:} The benchmark should provide a fair assessment for compatible systems. In particular, bias towards any one system should be limited.

  \item\label{req:easeofuse} \textit{Ease of Use:} The benchmark should be easy to configure and use with any compatible system.

  \item\label{req:clarity} \textit{Clarity:} The benchmark should present results to the user in a way that is suitable for system performance variability.

  \item\label{req:portable} \textit{Portability:} The benchmark should support common deployment environments and be easy to port to others.

  \item\label{req:scalability} \textit{Scalability:} Benchmark workloads should be scalable to accommodate benchmarking on increasingly powerful hardware or with more performant systems.
\end{enumerate}

\begin{figure}[t]
  \includegraphics[width=.9\linewidth]{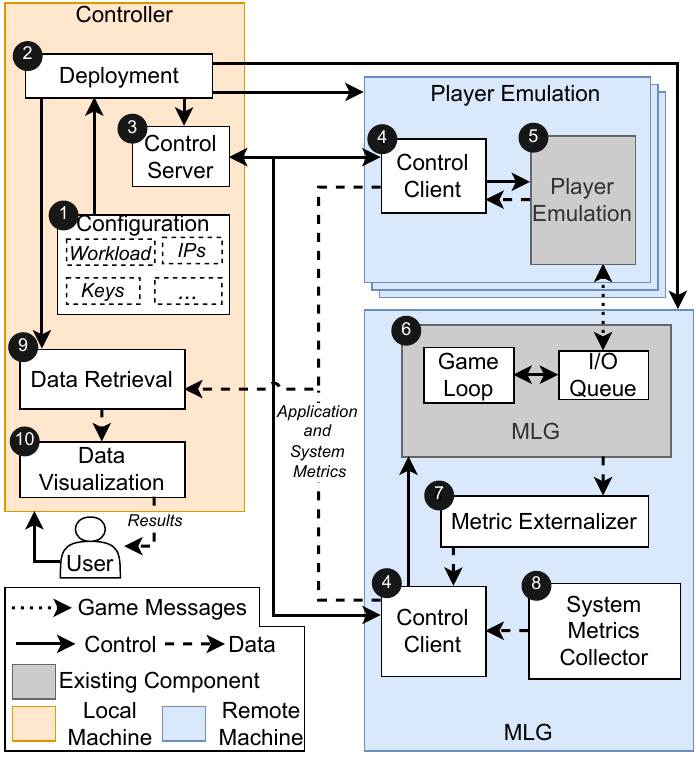}
  \vcutM{}
  \caption{Architecture of \toolname. Component \designref{6} is system under test. Component \designref{5} adapts tool~\cite{yardstick}.}
  \label{fig:benchmark}
  \vcutL{}
\end{figure}

\subsection{Design Overview}
Here we present the design of \toolname, our system for benchmarking of performance variability in \mves.
Figure~\ref{fig:benchmark} presents \toolname's high-level design.
We discuss the benchmark workloads~(addresses \textbf{R2}) and metrics~(partially addresses~\textbf{R3}) in more detail in~\Cref{sec:design:workloads} and~\Cref{sec:design:metrics} respectively.




In our design, the user mainly interacts with \toolname through its Configuration component~(\designref{1}).
The Configuration allows the user to capture performance variability by specifying the duration and number of iterations of experiments~(partially addresses~\ref{req:capture-variability}).
The Configuration further allows users to configure benchmark parameters, such as the systems under test and workload, and deployment parameters, such as machine IP addresses~(partially addresses~\ref{req:easeofuse}).

After specifying the configuration, the user launches \toolname.
This triggers the Deployment component~(\designref{2}), which deploys components and software dependencies to remote machines specified in the configuration.
This only requires the user needs to specify a set of IP addresses of \texttt{ssh}-accessible machines.
This makes \toolname portable~(\ref{req:portable}),
and allows users to evaluate \mve performance variability under cloud or self-hosted deployments.

When deployment is complete, the Deployment component hands control to the Control Server~(\designref{3}).
\toolname follows a Controller/Worker pattern,
with the Control Server as the controller, and the Control Clients as the workers~(\designref{4}).
The Control Server contains the operation logic, and is responsible for synchronizes the operation of all workers by exchanging messages with each Control Client running on each worker. The Control Server and Clients exchange the messages enumerated in Table~\ref{tab:cont_messages}.
Depending on the configuration, the Control Client runs either \emph{player emulation} or the \mve.

\toolname uses one or more workers for player emulation~(\designref{5}).
These workers emulate players by connecting the \mve and automatically sending player actions based on programmed behavior. \toolname implements this by using the player emulation from Yardstick~\cite{yardstick}, an existing \mve benchmark which we compare to \toolname in detail in~\Cref{sec:related-work}.

One worker runs the \mve~(\designref{6}), which is the system under test.
\toolname captures the \mve's performance variability metrics using the player emulator (\designref{5}), the metric externalizer~(\designref{7}), and the system metrics collector~(\designref{8}).
\Cref{sec:design:metrics} details
the operation of these components and the metrics they collect, including our novel metric to capture performance variability. 

When the benchmark experiments are done, the Control Server activates the Data Retrieval and Data Visualization components~(\designref{9} and \designref{10}),
This component moves the collected data from the worker nodes to the user's local machine, where it pre-processes the data through aggregation and reformatting.
The Data Visualization component~(\designref{10}) takes as input the processed data and automatically outputs 
producing basic plots for \mve performance and performance variability.
Users can view these plots, and, if desired, provide their own advanced plotting scripts for in-depth analysis
~(concludes~\ref{req:easeofuse},~\ref{req:clarity}).
\begin{table}[t]
  \begin{center}
    \caption{List of controller messages. Dest specifies what nodes the message is for where Y is player emulation, M is the server node and C is the controller.
    \label{tab:cont_messages}}
    \begin{tabular}{l l c}
      \hline
      \textbf{Message} & \textbf{Effect} & \textbf{Dest}\\
      \hline
       set\_server:\textit{server} & Specifies name of server & Y/M\\
       set\_jmx:\textit{jmx url} & Specifies JMX URL & M\\
       iter:\textit{iteration} & Specifies what iteration to start at & Y/M \\
       initialize & Starts the selected server & M\\
       log\_start & Starts metric logging tools & M\\
       log\_stop & Stops metric logging tools & M\\
       stop\_server & Stops running server & M\\
       connect & Starts player emulation & Y\\
       convert & Converts metric bin files to CSV & Y \\
       ok & Acknowledges the previous message & C\\
       keep\_alive & No-op, keeps TCP connection open & M/Y\\
       err:\textit{error} & Previous message has caused error & C\\
       exit & Stops the controller client & M/Y\\
      \hline
    \end{tabular}
  \end{center}
\end{table}


\subsection[Benchmark Workloads]{Benchmark Workloads
  (address \ref{req:valid-workloads},~\ref{req:fairness},~\ref{req:scalability})
}\label{sec:benchmark_workloads}
\label{sec:design:workloads}
This section presents \toolname's workloads.
\toolname uses the workload model presented in \Cref{sec:model},
which divides workloads in three main components: \emph{players}, \emph{terrain simulation}, and \emph{entities}.
By using this model, \toolname's workloads are applicable to \mves in general, thus avoiding favoring specific systems (partially addresses~\ref{req:fairness}).
In practice, the user specifies in the Configuration~(\designref{1}) only the \emph{player} and \emph{terrain simulation} parts of the workload,
as \emph{entities} are a result of terrain simulation (spawning, see \Cref{sec:Workloads}).

As future systems may become sufficiently performant to mitigate the impact of \toolname's workloads, \toolname supports workload scaling~(\ref{req:scalability}).
To increase \toolname's workload complexity, the user can specify an increased number of players to scale the player workload, and use \toolname's \emph{scale} parameter to select higher-complexity versions of the pre-configured workloads.

While \toolname supports arbitrary valid Minecraft worlds as workloads, the remainder of this section describes the workloads we design for use in our experiments to highlight performance variability based off our workload model and observable community use~(\ref{req:valid-workloads}).

The conceptual challenge of designing the benchmark workloads stems from the vast design space. \mves give players fine-grained control over the virtual environment,
resulting in an endless number of possible world permutations and a large variety in types of simulated constructs.

Additionally, finding evidence to support our selection proved to be challenging, as no peer-reviewed analysis of such artifacts currently exists and game operators do not want to share such information; searching for trustworthy communities and identifying suitable artifacts poses additional challenges. We detail our workload design and the evidence supporting it throughout this section.

\subsubsection{The Environment-Based Workloads}\label{sec:environment_workloads}
The environment workload is determined by the \mve's terrain generation and the terrain modifications made by players.
To obtain representative worlds and workloads, we reconstruct highly popular creations (templates of useful simulated constructs, see~\Cref{sec:model:workloads}) available on common sharing platforms in the \mve community.
Because the \mve community thrives on sharing player-created content with other players, this approach captures essential features of how the community uses these systems.

To cover the range of valid workloads~(\ref{req:valid-workloads}),
we include two worlds that result in a best-case workload and worst-case workload respectively.
During all \envworkloads, \toolname connects to the game a single player that performs no actions.
This is necessary to correctly capture the response time metric discussed in \Cref{sec:metrics}.
The remainder of this section describes the worlds and their resulting workloads.
We list the worlds used in Table~\ref{tab:worlds}.

\begin{table}[t]
  \begin{center}
    \caption{Minecraft worlds used as workload starting points by the \toolname benchmark.}
    \label{tab:worlds}
    \begin{tabular}{l l r}
      \toprule
      \textbf{Name} & \textbf{Properties}                      & \textbf{Size [MB]} \\
      \midrule
      Control       & Freshly generated world                  & 5.4                \\
      TNT           & Entity actions, terrain updates          & 6.3                \\
      Farm          & Resource Farm constructs                 & 26.0               \\
      Lag           & Complex simulated construct, stress test & 4.7                \\
      \bottomrule
    \end{tabular}
  \end{center}
  \vcutL{}
\end{table}

\begin{table}[t]
  \begin{center}
    \caption{Simulated constructs in Farm world and their author. Popularity measured in millions of views.
    \label{tab:farmconstructs}}
    \begin{tabular}{l r l c} 
      \toprule
      \textbf{Name} & \textbf{Amount}& \textbf{Author}& \textbf{Popularity}\\
      & & & [$10^6$ views] \\
      \midrule
      Entity Farm &12&gnembon~\cite{ytmobfarm} & 1.7\\
      Stone Farm &4&Shulkercraft~\cite{ytcobblefarm} & 1.3\\
      Kelp Farm &4&Mumbo Jumbo~\cite{ytkelpfarm} & 2.5 \\
      Item Sorter &1&Mysticat~\cite{ytitemsort}& 0.8 \\
      \bottomrule
    \end{tabular}
  \end{center}
\end{table}


The Control world results in a best-case workload while still being realistic.
The Control world is an unmodified world 
generated by Minecraft version 1.16.4 using the seed -392114485.
The measured results of this workload are used as a workload-level baseline to compare the other workloads.

The TNT world contains a 16-by-16-by-14 cuboid filled with TNT blocks which are set to explode around 20 seconds after a player connects.
In the systems tested, TNT operates by spawning an entity, which can be interacted with by other entities, including other TNT entities. Thus, when a large section of TNT is activated, the \mve must perform a large number of both entity-collision and physics calculations.
Intentionally creating large-scale TNT chain reactions is a popular activity, which can be observed in a plethora of community-made content. For example, a video from 2018 that shows a chain reaction of thousands of TNT blocks has 21~million views~\cite{yttnt}.

The Farm world contains multiple \emph{resource farms}, which are simulated constructs built by players to automatically generate in-game resources. The specific designs of the simulated constructs in this workload were sourced from popular community creators and each have 1.6~million views on average.
Table~\ref{tab:farmconstructs} gives an overview of all resource farms present in the Farm world, their respective authors and popularity.
 The Farm world has many different resource farm simulated constructs, which we list in Table~\ref{tab:farmconstructs}.
The Entity and Stone farms are 
activated at a fixed interval of around 4 seconds, whereas the Kelp and Item sorter use event-based activation. 
These farms rely on entities in their functioning, through spawning driven entities and manipulating their pathfinding, or through the creation of passive entities to represent item resources. A core feature of \mves is collection of resources from the game environment. The ability for players to construct simulated constructs that automate this process is both an intended and common behavior.
The Entity farm relies on spawning driven entities and manipulating their movement. The Stone and Kelp farm continuously destroy blocks, which create passive entities to represent items. These item entities are then transported through terrain simulation rules (e.g. liquid physics simulation).

The Lag world results in a worst-case workload.
This world contains a simulated construct known in the \mve community as a \emph{Lag Machine}.
Lag Machines are a specific subset of simulated constructs that are designed to cause high computational load for the \mve, either for the purpose of stress testing it, or to cause it to crash as part of a denial of service attack.
The design of the Lag Machine used in this workload is publicly available and provided by a community-creator with 52 thousand subscribers~\cite{ytlag}. It is chosen as it operates based on terrain simulation rules. Specifically, it uses many logic-gate constructs in a small area to cause a high volume of simulation rule activations. Importantly, the simulation rules this Lag Machine uses are generally non-malicious, and are used in many resource farm constructs, as well as forming the basis for simulated constructs such as an operational digital Computer~\cite{ytcomputer}.

\subsection{Configuration Parameters} \label{sec:Configuration_Parameters}

\toolname is highly configurable in order to be extensible to new Minecraft-like games, environments, workloads, and experiments~(\textbf{R5}). A list of configuration parameters are given in Table~\ref{tab:parameters}.

\begin{table}[t]
  \begin{center}
    \caption{List of configurable parameters. V, F and P refer to Vanilla, Forge and PaperMC respectively.
    \label{tab:parameters}}
    \begin{tabular}{l l l}
      \hline
      \textbf{Parameter} & \textbf{Specifies}& \textbf{Typical Value}\\
      \hline
      IPs  &  Nodes used & none\\
      SSL Keys  & Authentication & none \\
      Servers & Minecraft-like games & V, F, P\\
      World & World workloads & Control \\
      File Locations & Output directories & /ys/, /mc/\\
      Resume  & Continue experiment & False\\ 
      Ports& Network config & 25555/25565\\
      JMX URLs & Metric collection&Various\\
      JMX Ports&  Metric collection &25585-25635\\
      RAM & RAM (JVM argument) &4GB\\
      Affinity & CPU Affinity & 0xFFFFFFFF\\
      Number of Bots & Player count&25 \\
      Behavior & Player actions&Bounded random\\
      Duration  & Iteration length & 60 seconds\\
      Iterations  & Iteration amount & 1\\
      Scale & Workload intensity & 1\\ 
      \hline
    \end{tabular}
  \end{center}
\end{table}


The experiment can be configured by specifying the duration, number of iterations, what servers to run, how many players to connect and the behavior of those players. Since the IPs of the nodes that the deployment tool connects to can be anywhere that is network accessible, nodes can be chosen that are geographically distant or within the same datacenter in order to measure (or avoid) the performance impact caused by public network infrastructure.


\subsubsection{The Player-Based Workload}\label{sec:player_workloads}
\toolname uses a player-based workload facilitated by the player emulation component~(\designref{5}).
In this workload, \toolname is configured to connect 25 players which move randomly in a 32-by-32 area. The existing Yardstick benchmark~\cite{yardstick} focuses solely on the impact of player workload. So, we include this player workload to represent a high-density area in \mves and allow \toolname to compare the impact of \envworkloads with a traditional player-based workload.
We select a player count of 25 based on the Minecraft Wiki's dedicated server recommendation~\cite{serverrequirements}, as well as the recommendations from various commercial cloud providers (see \Cref{sec:experimental_setup}).

\subsection[Metrics]{Metrics
  (address \ref{req:capture-variability},~\ref{req:extensive-exps-and-metrics},~\ref{req:fairness})}
\label{sec:metrics}
\label{sec:design:metrics}

This section describes the \begin{addtext}application-level and system-level metrics \end{addtext} metrics collected by \toolname, selected to fulfill \ref{req:extensive-exps-and-metrics}.
In ~\Cref{sec:metric} we describe our novel \emph{\newmetric} which quantifies performance variability ~(concludes~\ref{req:capture-variability}).
Our \newmetric metric and all application and system metrics are general to \mves to avoid bias for specific implementations (concludes~\ref{req:fairness}).
Table~\ref{tab:metrics} gives an overview of the collected metrics.

\begin{table}[t]
  \begin{center}
    \caption{Metrics collected by \toolname. The metric type is \underline{D}erived, \underline{A}pplication level, or \underline{S}ystem level.}
    \label{tab:metrics}
    \vcutS{}
    \begin{tabular}{l l l r}
      \toprule
      \textbf{Type} & \textbf{Metric}   & \textbf{Description}                          \\
      \midrule
      D             & \newmetricfull    & Tick instability (see~\Cref{sec:metric}) \\
      \midrule
      A             & Response time     & Round trip latency for clients                \\
      A             & Tick duration     & Duration of each tick                         \\
      A             & Tick distribution & Tick time by workload                         \\
      \midrule
      S             & CPU               & CPU utilization                               \\
      S             & Memory            & Memory usage                                  \\
      S             & Threads           & Thread total                                  \\
      S             & Disk I/O          & Bytes read/written                            \\
      S             & Network I/O       & Bytes sent/received                           \\
      \bottomrule
    \end{tabular}
  \end{center}
  \vcutL{}
\end{table}

\subsubsection{Application-level Metrics}
\label{sec:design:metrics:application}

\toolname collects three application level metrics:
\emph{response time}, \emph{tick duration}, and \emph{tick distribution}.

Response time is how system latency becomes visible to the user.
Lower values are better, and we use existing latency thresholds for the game becoming noticeable and unplayable at 60ms and 116ms respectively~\cite{latency60ms,unplayable118ms}.

The response time is measured as the time between a player taking an action and the results of that action becoming visible.
During this time, the action is sent over network to the \mve server, added to an input queue, simulated during the next tick, has its resulting state changes added to the output queue, and then sent back to the client over the network. In our workload model (Figure~\ref{fig:operational}), this is visible as the time difference between the client sending player actions (\designref{A}) and receiving state updates (\designref{C}).
\toolname captures this metric by having a player send \emph{chat} messages to all players (including itself), and measuring how long it takes for the player to receive its own message.

While tick duration and tick distribution cannot be directly observed by players,
\mves typically expose these metrics through interfaces commonly used by debugging tools.
\toolname's Metric Externalizer~(\designref{7}) uses these interfaces to gain access to these metrics without requiring access to the game's source code.
As a consequence, \toolname is easily configured to work with new \mves.

The \emph{tick duration} is the amount of time it takes the \mve to complete a single iteration of the game loop,
and \emph{tick distribution} is the percent of tick time the \mve spent simulating each workload component, such as simulating entities.
Both metrics are directly related to game response time and are important indicators of game performance.
More detail about the relationship between these metrics is available in \Cref{sec:model:mve-archi}.



\subsubsection{System-level Metrics}
\label{sec:design:metrics:system}

\toolname captures system-level metrics to allow users to perform a more in-depth performance analysis.
\toolname collects system-level metrics using the System Metrics Collector~(\designref{8}),
which queries the operating system twice per second.

\toolname collects CPU utilization, memory usage, the number of operating-system threads associated with the \mve, disk I/O, and network I/O.
These metrics allow users to analyze causes of high tick duration,
and check for potential performance bottlenecks.


\section{\newmetricfull metric}
\label{sec:metric}

In this section we present our novel \newmetricfull metric.
We present a definition, analyze its properties, and compare it to existing metrics.

\subsection{\newmetricfull Definition}
\label{sec:design:metrics:derived}
\label{sec:design:metrics:jitter}
\label{sec:metric:definition}

In the context of online gaming, players prefer stable performance to unstable, but on average faster, performance~\cite{DBLP:journals/cacm/ChenHL06, ries2008empirical, normoyle2014player}. Stability facilitates predictability, allowing players to acclimate to game update delays, up to a point.
Thus, it is beneficial to quantitatively analyze the stability of game performance by analyzing the variability of game cycles, or ticks~(see~\Cref{sec:model:mve-archi}). However, existing measures of variability are insufficient, because they do not capture the order of
ticks, outlier values, the duration of the trace, or a combination of these elements.

We describe here our novel \newmetricfull~(\newmetric), a normalized metric
based on \emph{cycle-to-cycle jitter}~\cite{jitterpll,JitterPhase}. 
In the context of \mves, we measure cycle-to-cycle jitter as the difference in delay between consecutive ticks~(see~\Cref{sec:model:mve-archi}).
This delay starts to vary when the game becomes overloaded. We compute the \newmetric as the normalized sum of \mve cycle-to-cycle jitter.
The cycle-to-cycle jitter considers only the
difference between two consecutive ticks; reports for this metric include the maximum or moving average value. Novel in this
work, our \newmetric metric sums the differences, and normalizes the result\remove{ing sum by twice the trace duration}.

The full metric equation is shown in Equation~\ref{eqn:jitter}, where
$N_e$ is the expected number of ticks,
$t_i$ is the duration of the~$i^{\text{th}}$ game tick,
$b$ is the delay between ticks when the \mve runs at its intended frequency,
$\max ( b, t_i )$ is the period of tick $i$, and
$N_a$ is the actual number of ticks.

When a tick lasts longer than the $b$ value, the proceeding tick is delayed. Thus, if the game meets its performance requirements $N_a = N_e$, but if it becomes overloaded $N_a \leq N_e$~(i.e., $\exists i : t_i > b \implies N_a \leq N_e$).


\begin{equation}\label{eqn:jitter}
  \text{ISR} = \frac{\sum_{i=1}^{N_a} \left| \max(b, t_i) - \max(b, t_{i-1}) \right| }{N_e \times 2b}
\end{equation}

Using this metric as a measure of variability, the range of possible values is 0 to 1. A \newmetric of 0 indicates no variability:
the tick period is constant for all ticks in the trace.
A \newmetric of 1 indicates maximum variability, and is reached when the sum of differences between consecutive ticks is equal to twice the duration of the trace~(i.e., $N_e \times 2b$).
This value is reached when tick periods alternate between their intended value and extremely large values.

\begin{figure}[t]
  \centering
  \begin{subfigure}{0.48\linewidth}
    \includegraphics[width=\linewidth]{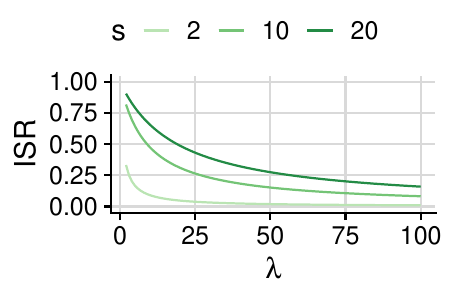}
    \caption{Behavior of \newmetric for varying outlier periods ($\lambda$).}
    \label{fig:metric:sr-over-period}
  \end{subfigure}
  \hfill
  \begin{subfigure}{0.48\linewidth}
    \includegraphics[width=\linewidth]{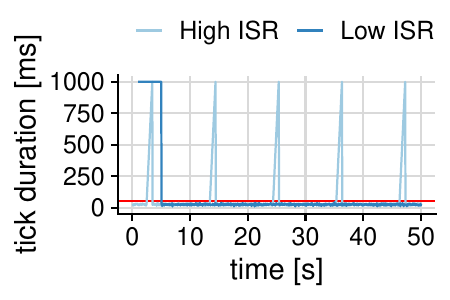}
    \caption{Example traces resulting in different \newmetric values.}
    \label{fig:metric:sr-example}
  \end{subfigure}
  \caption{Numerical analysis of \newmetricfull. Higher values indicate higher performance variability. $s$ indicates the outlier scaling factor. $\lambda$ indicates the period between outliers in number of ticks.}
  \label{fig:metric:sr}
  \vcutL{}
\end{figure}

\subsection{Analysis of \newmetric Behavior}
\label{sec:metric:numanalysis}

We analyze the behavior of \newmetric by modeling a trace where every $\lambda$~ticks, one tick has a duration of $sb$, while the others all have duration $b$. This means 1 in $\lambda$ ticks exceeds the performance requirement by a factor $s$. This allows expressing \newmetric as $\newmetric=\frac{s-1}{s+\lambda-1}$.
A plot of this function and an example trace based on this model are available in Figure~\ref{fig:metric:sr}.

Figure~\ref{fig:metric:sr-over-period} shows how \newmetric responds to outlier scaling and frequency.
The horizontal axis shows~$\lambda$, which is the number of ticks between outliers, and the vertical axis shows the value of \newmetric. The curves show the value of \newmetric for three values of $s$. The three values of s~(2, 10, 20) indicate that all outliers exceed the latency requirement by a factor 2, 10, or 20, respectively.

\emph{The plot shows that \newmetric increases when outliers become larger~(increasing s),
and when outliers are more common~(lower $\delta$).}
For example, a tick exceeding $b$ by a factor 10~($s=10$) every 25 ticks~($\lambda = 25$) results in an ISR value of 0.26.


Figure~\ref{fig:metric:sr-example} shows two example traces: High ISR and Low ISR.
Both traces contain 1000 ticks.
The horizontal axis shows time, and the vertical axis shows tick duration.
Most ticks have a duration below 50\,ms (b), but each trace has five outliers with a scaling factor of 20, resulting in a 1\,second spike.
For the Low ISR trace, all outliers are the start of the trace,
whereas in the High ISR trace the outliers are evenly distributed over time.
While the statistical distributions of the traces are identical,
the ISR for the Low ISR trace is 0.009, the ISR for the High ISR trace is 0.15, an order of magnitude higher.

\begin{table}[t]
  \centering
  \caption{Comparison of \newmetric with existing variability metrics.}
  \label{tab:metric_comp}
  \vcutS{}
  \begin{tabular}{lccc}
    \toprule
    \textbf{Metric}                         & \textbf{Order}     & \textbf{Irregular} & \textbf{Normalized} \\
                                            & \textbf{Dependent} & \textbf{Sampling}  &                     \\
    \midrule
    Standard deviation                      & \xmark             & \xmark             & \xmark              \\
    Allan variance~\cite{riley2008handbook} & \cmark             & \xmark             & \xmark              \\
    Jitter~\cite{schulzrinne2003rfc3550}    & \cmark             & \cmark             & \xmark              \\
    \textbf{\newmetric}                     & \cmark             & \cmark             & \cmark              \\
    \bottomrule
  \end{tabular}
\end{table}

\subsection{Comparing \newmetric to Alternative Metrics}
\label{sec:metric:compare}

Table~\ref{tab:metric_comp} compares \newmetric to existing measures of variability.
Standard deviation captures spread from an average value. It is not order dependent, and thus is not a measure of stability, but dispersion.

Allan variance is used in the field of electrical engineering as a time domain measure of frequency stability, typically applied to clocks or oscillators~\cite{riley2008handbook}. Allan variance is order dependent, but relies on a constant sampling frequency and a continuous sampling domain. Neither property is applicable to the duration of tick values.

Jitter is defined in the domain of networking as the smoothed absolute difference between consecutive packets~\cite{schulzrinne2003rfc3550}. While most similar to \newmetric, which is based on cycle-to-cycle jitter, it is not normalized, but rather reported as an average and defined for any packet, rather than an entire sampling duration.

\section{Real-World Experiments}\label{sec:experiments}

To address contribution~\ref{contrib:experiments}, we present here the setup and results from our real-world experiments.
We use \toolname to evaluate the performance variability of three \mves: Minecraft, Forge, and PaperMC. Our experiments use the workloads described in \Cref{sec:design:workloads}, and are conducted on two commercial cloud environments, Amazon~AWS and Microsoft~Azure, as well as on DAS-5, a supercomputer for scientific workloads~\cite{das5}. 


\label{sec:main_findings}
\begin{enumerate}[label=\textbf{MF\arabic*}]
  \item\label{mf:variability-player-experience} Performance variability can make
        \mves unplayable~(\Cref{sec:mf1}).
        We find that the maximum response time can be up to 20.7 times higher than the arithmetic mean, and exceed by more than a factor of~7.4 the threshold for playable games.

  \item\label{mf:envworkloads-perf-variability} \Envworkloads cause significant performance variability~(\Cref{sec:mf2}).
        We find that \envworkloads introduce significant performance variability, increasing \newmetric by 0.04 up to 0.92. This variability can overload popular \mves by 58 times the normal tick duration and even crash the game.


  \item\label{mf:cloud-perf-variability}  \mves exhibit increased variability in commercial cloud environments compared to self-hosted environments~(\Cref{sec:mf3}).
        We show that both clouds, AWS and Azure, introduce additional performance variability between iterations of the same workload compared to the local environment, \das. The choice of cloud causes a 1.39 up to 15.44 times increase in \newmetric IQR and a 1.09 up to 5.61 times increase tick time IQR. The minimum observed \newmetric for both clouds exceeding the maximum observed \newmetric on \das.

  \item\label{mf:entities-expensive} Processing the state of entities is computationally expensive~(\Cref{sec:mf4}). Entity-related computations account for a majority of the non-idle tick distribution in all experiments.

  \item\label{mf:default-hardware-insufficient} The common hardware resource recommendations are insufficient to avoid performance variability~(\Cref{sec:mf5}).
        The recommended node size exhibits high performance variability and high mean tick duration.
        Larger node sizes result in lower values of both, such that a node with 8~vCPUs reduces performance variability and mean tick duration to acceptable levels.


\end{enumerate}

\subsection{Experimental Setup}\label{sec:experimental_setup}

In this section we describe our experimental setup.
In our experiments, we evaluate three \mves, i.e., the systems under test, in three different environments.


\subsubsection{System under test}

We use in our experiments three \mves that use the Minecraft protocol: the original Minecraft as developed by Mojang~\cite{mcserver}, \textit{Forge},
and \textit{PaperMC}.
We select these services because of their popularity and utility.

We choose the official Minecraft server~\cite{mcserver}, as it is the default service that most users and hosts operating Minecraft servers employ. This includes users of the server hosting service \textit{Realms}, which is advertised inside the Minecraft client~\cite{Realms}. Notably, the Minecraft service does not allow for gameplay modifications~(``\textit{mods}") of any kind.

\emph{Forge} is the most popular \mve for operating modified (i.e., \emph{modded}) services~\cite{forge}. Of the top-50 most downloaded Minecraft mods,
45~work exclusively with Forge.
Of the 5~mods that are not exclusive to Forge, only one is incompatible with it~\cite{mcmods}.
\emph{PaperMC} is marketed as a high-performance alternative to Minecraft~\cite{papermc}.
While the PaperMC project does not quantify its performance improvement over Minecraft, it does provide documentation of its optimizations, which include extensive changes to threading models and virtual environment processing.

\subsubsection{Deployment Environment}

\begin{table}[t]
  \begin{center}
    \caption{Hardware recommendations from companies that offer cloud hosting of Minecraft-like games. Field marked NP are information not provided to consumers, fields marked V are variable. \label{tab:hardware_recs}}
    \begin{tabular}{l c c c}
      \hline
      \textbf{Service} & \textbf{RAM [GB]} & \textbf{vCPU[\#]} & \textbf{CPU Speed[GHz]} \\
      \hline
      \href{https://www.hostinger.com/minecraft-server-hosting}{Hostinger}~\cite{hostinger} & 3 & 3 & NP \\
      \href{https://server.pro/}{Server.pro}~\cite{server.pro} & 4 & 2 & 2.4 \\
      \href{https://www.skynode.pro/}{Skynode}~\cite{skynode} & 4 & 2 & 3.6\\
      \href{https://scalacube.com/hosting/server/minecraft/plans/public?}{ScalaCube}~\cite{scalacube} & 3 & 2 & 3.4 \\
        
      \href{https://nodecraft.com/}{Nodecraft}~\cite{nodecraft} & 4 & NP & 3.8\\
      \href{https://apexminecrafthosting.com/}{Apex Hosting}~\cite{apex} & 4 & NP & 3.9\\
      \href{https://ggservers.com/}{GGServers}~\cite{ggservers} & 4 & NP & 3.2\\
      \href{https://www.bisecthosting.com/}{BisectHosting}~\cite{bisect} & 4 & NP & 3.4\\
      \href{https://shockbyte.com/}{Shockbyte}~\cite{shockbyte} & 4 & NP & 4.0\\
      \href{https://cubedhost.com/}{CubedHost}~\cite{cubedhost} & 2.5 & NP & 4.5\\
      \href{https://serverminer.com/buy-minecraft-server}{ServerMiner}~\cite{server} & 3 & NP & 4.0\\
      \href{https://www.akliz.net/pricing}{Akliz}~\cite{akliz} & 4 & NP & 3.4\\
      \href{https://ramshard.com/hosting/minecraft}{RamShard}~\cite{ramshard} & 2 & NP & 4.0\\
    \href{https://mcprohosting.com/}{MCProHosting}~\cite{mcpro} & 2 & NP & NP\\
      \href{https://www.gtxgaming.co.uk/}{GTXGaming}~\cite{gtxgaming} & 3 & NP &3.8 \\
    \href{https://stickypiston.co}{StickyPiston}~\cite{stickypiston} & 2.5 & NP & NP\\
      \href{https://hosthavoc.com/minecraft}{HostHavoc}~\cite{hosthavoc} & 4 & NP &4\\
    \href{https://feroxhosting.nl/}{Ferox Hosting}~\cite{ferox} & 4 & NP &NP\\
      \href{https://aquatis.host/minecraft-hosting/}{Aquatis}~\cite{aquatis} & 4 & NP & 4.2\\
      \href{https://pebblehost.com}{PebbleHost}~\cite{pebble} & 3 & NP & 3.7\\
      \href{https://www.meloncube.net/}{MelonCube}~\cite{meloncube} & 4 & NP & 3.4\\
      \hline
      \href{https://docs.microsoft.com/en-us/gaming/azure/reference-architectures/multiplayer-basic-game-server-hosting}{Azure}~\cite{azureserver} &  4 & 2 & V\\
      \href{https://web.archive.org/web/20201126074839/https://aws.amazon.com/getting-started/hands-on/run-your-own-minecraft-server/}{AWS}~\cite{amazonserver} & 1 & 1 & V\\
      \hline
    \end{tabular}
  \end{center}
\end{table}

We evaluate the \mves in two commercial cloud environments, Amazon~AWS and Microsoft~Azure,
and \das, a supercomputer for academic and educational use~\cite{das5}.
We choose AWS and Azure because they are the two cloud environments with the biggest market share, with 32\% and 20\% respectively~\cite{cloudproviders}.
We use \das to evaluate how commercial cloud environments affect the performance variability of \mves, compared to self-hosting these games on dedicated hardware.


Our experiments on cloud environments use \emph{T3.Large} and \emph{Standard\_D2\_v3} nodes respectively.
Both node types are equipped with 2\,vCPUs and 8\,GB memory. We choose these nodes based on the default hardware configurations recommended by Minecraft service providers as well as guidelines published by AWS and Azure~\cite{azureserver,amazonserver}.


On \das, we use a regular node, which is equipped with a dual 8-core 2.4\,GHz processor and 64\,GB memory, and limit the number of CPU cores available to the \mve by setting its CPU affinity to two cores, unless indicated otherwise.
Because the \mves used in our experiments run on the Java Virtual Machine (JVM), we limit memory available to the \mve in both cloud an \das nodes by setting the JVM's maximum heap size to 4\,GB.

Table~\ref{tab:hardware_recs} lists a sampling of hardware recommendations from commercial cloud companies that offer Minecraft-like game hosting. If no plan was marked ``recommended," data is taken from plans that are comparable to recommended plan on other services. From these recommendations we find that~2\,vCPU and~4\,GB RAM is the most common configuration. On AWS and Azure it is possible to select specific configurations of hardware, however, to ensure that the Benchmark metric tools have sufficient memory during the experimental duration, we use nodes with at least~8\,GB RAM and limit the heap memory of the Minecraft-like game to~4\,GB using the \texttt{-Xmx} JVM argument in all experiments.



\subsection{\ref{mf:variability-player-experience}: Performance variability can make \mves unplayable}\label{sec:mf1}
\label{sec:experiments:variability-player-experience}

\begin{figure}[t]
    \includegraphics[width=\linewidth]{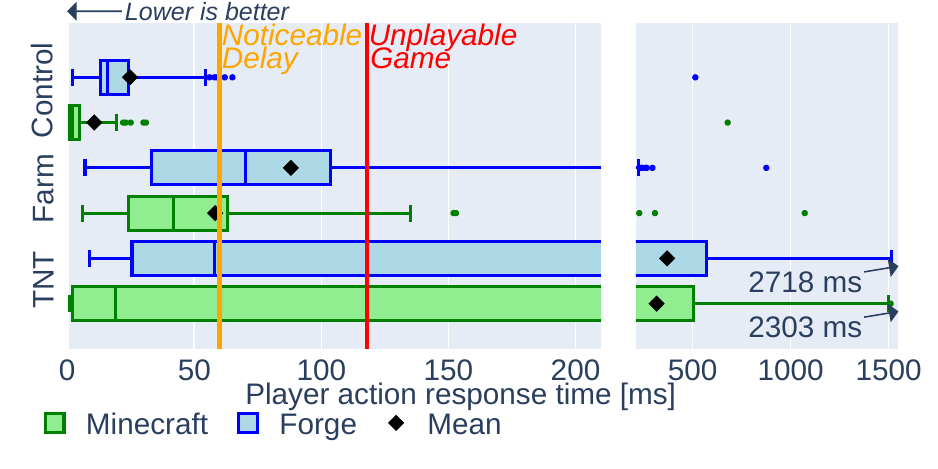}
    \vcutL{}
    \caption{Game response time in AWS environment when running separate \envworkloads.
        Whiskers indicate 5th and 95th percentile respectively.
        The black diamonds indicate arithmetic mean.}
    \label{fig:aws_response_time}
\end{figure}

Due to significant performance variability, the median and mean game response times give an optimistic view of game
performance, and is worse than the performance observed by players.
Figure~\ref{fig:aws_response_time} depicts the result, and shows that the 95th percentile of game response time can be up to 4.1 times higher than the arithmetic mean, and exceed by more than a factor of~12.8 the threshold that makes the game unplayable.
In real-time games, a temporary spike in delay can significantly affect the user's experience, similar to a temporary freeze in a phone call or video stream.

Figure~\ref{fig:aws_response_time} shows the response time (horizontal axis)
for two \mves (Minecraft in green, and Forge in blue) under three different workloads (vertical axis).
The workloads and response time metric are described in \Cref{sec:design:workloads} and \Cref{sec:design:metrics:application} respectively.
PaperMC is omitted as it uses a dedicated asynchronous thread for chat messages, separately from the game tick.
The whiskers extend to the 5th and 9th percentiles, respectively,

and the black diamond indicates the arithmetic mean.
The \emph{Noticeable Delay} line~(at~60\,ms, in orange) and \emph{Unplayable Game} line~(at~118\,ms, in red) indicate high game-latency which respectively marks the values where latency becomes noticeable to players and makes the game so unresponsive it becomes unplayable~\cite{latency60ms,unplayable118ms}.



Under the Control workload~(top two boxes), the 95th percentile is below the noticeable threshold for both Minecraft and Forge.
However, the maximum value for Forge~(514\,ms) is 20.7~times larger than the mean, and the maximum value for Minecraft~(679\,ms) is exceeds by 7.4~times the Unplayable threshold at 118\,ms.
These outliers occur directly after a player connects to the game.
This means that, even with good average performance, the game can still be unplayable if players frequently connect, which is a common occurrence in online multiplayer games.

Compared to the Control workload, the Farm and TNT workloads show significantly more performance variability, pointing to a further degradation of player experience.
\emph{In all cases, the mean and median values give an overly optimistic view of the game's performance.}
For the Farm workload, the mean and median values for Forge~(third bar from the top) indicate the response time is noticeable, but not unplayable. However, the plot shows a 95th percentile of 225.8\,ms, which is 1.9 times as high as the Unplayable threshold.
For Minecraft~(fourth bar from the top), the mean and median values indicate that the response time is not noticeable for players.
However, the plot shows that performance variability causes the response time to exceed the Noticeable threshold more than 25\% of the time~(box's right edge exceeds the Noticeable threshold), and exceeds the Unplayable threshold more than 5\% of the time.
The TNT workload causes the highest performance variability for both Forge and Minecraft (bottom two boxes, 547\,ms interquartile range for Forge and 503\,ms for Minecraft).
In both cases, the median response time is below the noticeable threshold, while the
95th percentiles are 12.7 times the unplayable threshold, and the maximum observed values (indicated with black arrows) are at least 19 times larger than the unplayable threshold.


From results in this section, we conclude that the mean and median values give an overly optimistic view of \mve performance, and that performance variability in \mves results in noticeable and unplayable game latency,
impacting players.


\begin{figure}[t]
    \includegraphics[width=\linewidth]{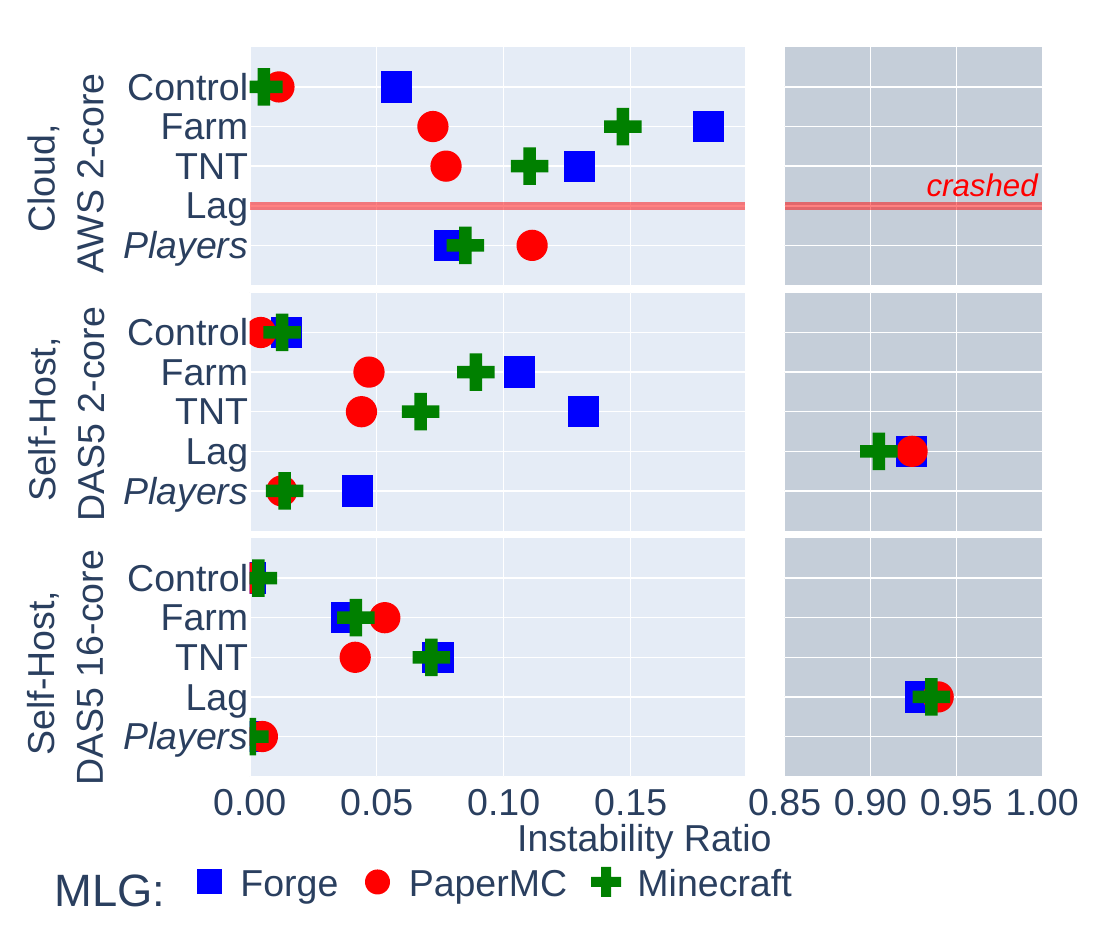}
    \vcutL{}
    \caption{\newmetric for each \mve on the AWS and \das environments. The Lag workload crashed all \mves on AWS; see text for explanation.
        \Cref{sec:player_workloads} defines the ``Players'' workload.}
    \label{fig:paw_jitter_scatter}
    \vcutM{}
\end{figure}

\subsection{\ref{mf:envworkloads-perf-variability}: \Envworkloads cause significant performance variability}\label{sec:mf2}
\label{sec:experiments:envworkloads-perf-variability}

\begin{figure}[t]
    \includegraphics[width=\linewidth]{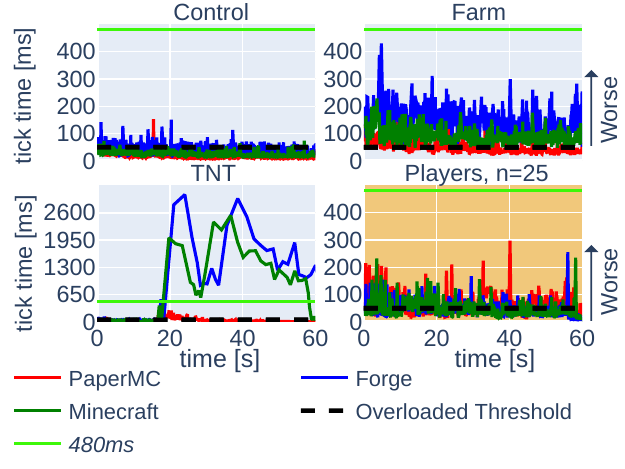}
    \vcutL{}
    \caption{Tick time over time for each \mve in the AWS environments running the Control, Farm, TNT and Players workloads. The Lag workload on AWS is omitted as each \mve crashes. \Cref{sec:player_workloads} defines the ``Players'' workload.}
    \label{fig:paw_tick_line}
\end{figure}

\Envworkloads cause significantly increased performance variability on each game and in each environment tested,
and can overload or crash the game.
Figure~\ref{fig:paw_jitter_scatter} shows the performance variability of each \mve when running \envworkloads on AWS and \das.
Compared to the control workload, each \mve on each environment exhibits higher performance variability when operating \envworkloads.

Figure~\ref{fig:paw_jitter_scatter} shows performance variability, quantified using \newmetric (see Equation~\ref{eqn:jitter}).
The three top-level rows show three environment configurations, each containing five workloads.
The color and shape of the marker indicate one of three \mves.


\Envworkloads (i.e., Farm, TNT, Lag) cause significantly higher performance variability than the player workload and control workload for all games in all environments, with the exception of PaperMC on AWS (red circles in the top row).
This provides evidence that \envworkloads cause significant performance variability.
Further analysis into the behavior of PaperMC reveals that it contains performance optimizations specifically for handling TNT explosions, improving its performance on the TNT workload, and Redstone, a simulated block type which is used in the Farm workload (analysis of PaperMC given in ).
This provides evidence that the performance variability caused by these \envworkloads are known to the \mve community and can (at least partially) be addressed through engineering.


Of all workloads, the Lag workload causes the most performance variability.
Further analysis reveals that this happens because this workload consists mainly of parts which are only simulated every other tick, causing the game to alternate between extremely short and extremely long ticks.
This maximizes the value of \newmetric, which is based on the difference in duration between consecutive ticks.
There are no results for running the Lag workload on AWS because all three \mves crash when a player joins and the environment simulation begins. The corresponding increase in tick duration causes the player's connection to time-out, forcing each \mve to stop. 


Figure~\ref{fig:paw_tick_line} shows the game's tick duration over time for each game when running on AWS. The dashed black line indicates the overloaded threshold at 50\,ms,
and the green line allows calibrating the vertical axis across the four sub-plots.




The stability observed when running the Control workload in Figure~\ref{fig:paw_jitter_scatter} is visible in the top-left sub-plot in Figure~\ref{fig:paw_tick_line} as three relatively stable curves with few spikes.
In contrast, the high performance variability observed for the Farm and TNT workloads is visible in the top-right and bottom-left sub-plots as jittery curves.
The Farm workload depicted in the top-right shows curves which change value at high frequency, resulting in high \newmetric.
PaperMC's tick durations are frequently below the 50\,ms threshold, resulting in lower \newmetric.
The TNT workload depicted in the bottom-left shows curves which change value at a much lower frequency, but reach significantly higher values, exceeding 2500\,ms for both Minecraft and Forge.
Similar to the Farm workload, PaperMC's tick durations are often below 50\,ms, resulting in lower \newmetric.

\subsection{\ref{mf:cloud-perf-variability}: \mves exhibit increased variability in commercial cloud environments}
\label{sec:mf3}
\label{sec:experiments:cloud-perf-variability}
\begin{figure}[t]
    \includegraphics[width=\linewidth]{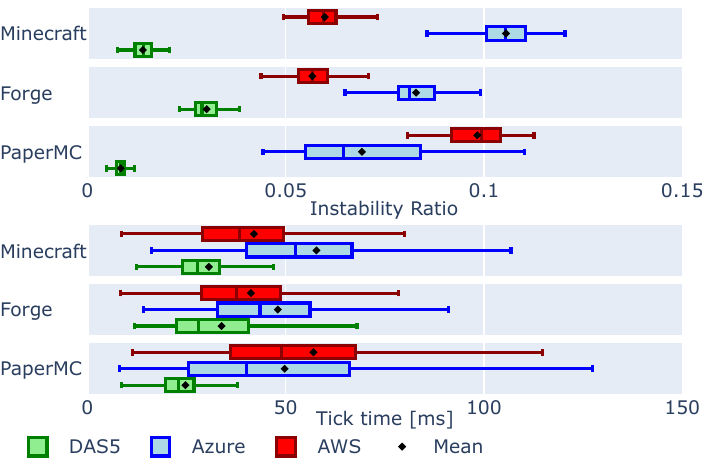}
    \vcutL{}
    \caption{Distribution of tick times and \newmetric from 50 iterations of the Players workload. Whiskers extend to $\pm$ 1.5 $\times$ IQR, bounded by the minimum and maximum values.} 
    \label{fig:50_iter_jitter}
    \vcutL{}
\end{figure}
In our experiments, all \mves show increased performance variability in terms of both variability (i.e. \newmetric) and tick times, when run on the AWS and Azure cloud environments, compared to the self-hosted \das.
Figure~\ref{fig:50_iter_jitter} shows \newmetric and tick time distribution across 50~iterations of the Player workload~(see \Cref{sec:player_workloads}) of all three games (on the vertical
axis) in \das (green), Azure (blue), and AWS (red).



The results show that all three games are the most stable, with the lowest median \newmetric (line inside boxes) and the lowest \newmetric overall, when run on \das.
The \emph{maximum} \newmetric observed on the \das is 0.021 (Forge), which is smaller than 0.029, the \emph{minimum} \newmetric observed in AWS and Azure~(PaperMC). Distribution of tick time follows a similar trend, with each game exhibiting lowest mean and median tick time on the \das, as well as the smallest interquartile range (IQR).

From this result, we highlight two surprising observations.
First, no game performs best in all environments.
On \das, PaperMC performs best, slightly outperforming Minecraft with a median \newmetric of 0.007 and 0.010 respectively.
Although PaperMC also has the lowest median \newmetric on Azure, it simultaneously has the highest IQR of both \newmetric, 0.028 compared to Forge's 0.009 and Minecraft's 0.011, and tick time, 40.75 to Forge's 23.25 and Minecraft's 26.71.
Moreover, on AWS, PaperMC is the worst performing game, with a median \newmetric of 0.094 and a median tick time of 48.98.
Second, neither cloud performs best for all games.
While AWS performs better for Minecraft and Forge, Azure performs best for PaperMC.


Increased performance variability in commercial cloud environments is a well-documented phenomenon~\cite{variability,tamingvariation,blackboxvariation,storagevariability}, with a wide variety of sources identified for the cause of increased variability, including hardware manufacturing differences, shared tenancy of hardware and networks, specific software configurations, and resource allocation and scheduling systems.
With so many variables operating in the context of commercial cloud hosting, it is infeasible to identify a single source responsible for the variability of these games, especially since commercial cloud hosting companies do not make internal data on resource allocation and shared tenancy publicly available.
However, we can conclude that this variability observably impacts the performance of \mves, and can be compared between \mves and commercial cloud services.

\subsection{\ref{mf:entities-expensive}: Processing entity-state is computationally expensive}\label{sec:mf4}
\label{sec:experiments:entities-expensive}

\begin{figure}[t]
    \includegraphics[width=\linewidth]{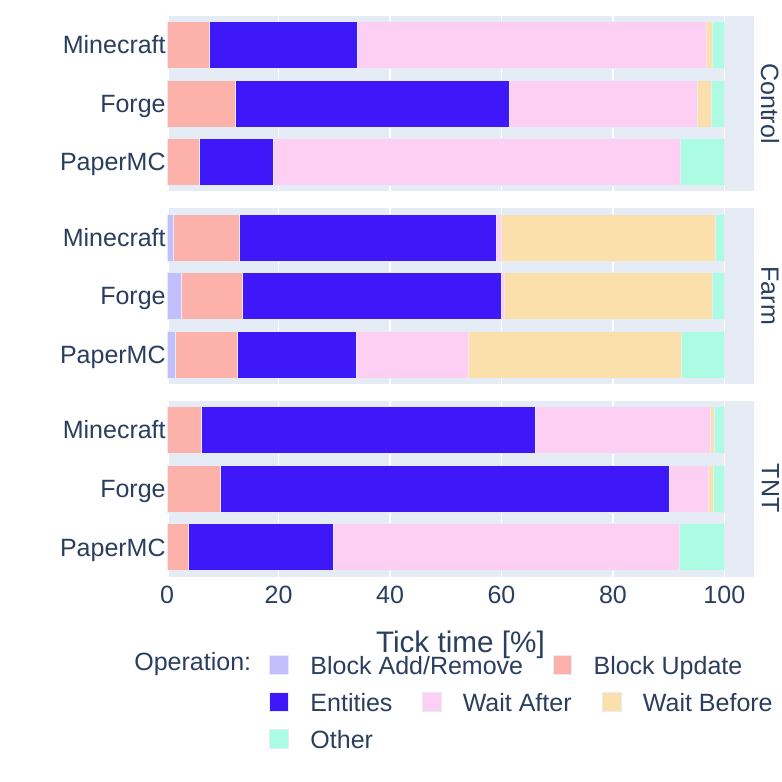}
    \caption{Distribution of tick duration attributed to specific \mve features on AWS environment. ``Other'' is the unspecified category.} 
    \label{fig:aws_tick_percents}
\end{figure}

Entity workload components account for a large majority of computation time and state update messages.

Figure~\ref{fig:aws_tick_percents} shows that entity related workload components contribute to a majority of non-waiting tick computation time. After entities, the next most time intensive component is environment rule processing, and then block creation or destruction. 

The horizontal axis is percentage of tick computation time. The vertical axis on the left is \mve and Workload on the right. The color of each bar indicates workload operation, and the width of each bar corresponds to their share of tick computation time throughout the experimental duration on the AWS environment. 

Entities account for a majority of non-waiting tick time during every workload on each server. Most notably, PaperMC has a much smaller proportion of entity calculation time under each workload compared to Minecraft and Forge. During the TNT workload, in which Minecraft and Forge show large percentages of entity tick computation time compared to both the Control and Farm workload, PaperMC has only a small increase. This reduction in Entity computation may explain how PaperMC manages its comparatively high performance during the TNT workload as seen in \ref{mf:variability-player-experience}.

\begin{table}[t]
  \begin{center}
    \caption{Percentage of network messages sent during experiments on AWS that are related to entities. Computation refers to percentage of messages sent and communication refers to percentage of bytes sent.
    \label{tab:messages}}
    \begin{tabular}{l l r r}
      \hline
      \textbf{Server} & \textbf{Workload} & \textbf{Computation} & \textbf{Communication}\\
      \hline
       & Control & 97.5& 3.8 \\
      Minecraft & Farm  & 91.7& 17.4\\
       & TNT & 97.0& 9.8 \\
       \hline
       & Control  & 97.2& 3.2\\
      Forge & Farm  & 86.7& 9.7\\
       & TNT  & 97.1& 10.3\\
       \hline
       & Control &89.1& 1.3 \\
      PaperMC & Farm  & 47.5& 1.2\\
       & TNT  &94.8& 3.5\\
      \hline
    \end{tabular}
  \end{center}
\end{table}

Table~\ref{tab:messages} shows that entity-related state updates account for a majority of messages sent to the client from the server in all configurations except PaperMC running the Farm workload. Conversely, entity-related state updates account for only a small percent of network bytes sent. 

Thus, we find that efficient computation and dissemination of entity state is a crucial performance challenge to \mves. Unlike environment processing, which exhibits spacial locality; entities require computation and take actions regardless of proximity to players, terrain or other entities. 
\subsection{\ref{mf:default-hardware-insufficient}: Using recommended hardware results in significant performance variability}
\label{sec:mf5}
\label{sec:experiments:default-hardware-insufficient}

\begin{figure}[t]
    \includegraphics[width=\linewidth]{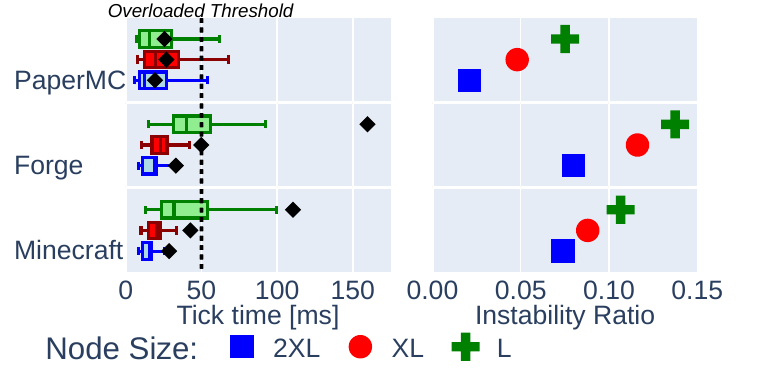}
    \vcutL{}
    \caption{Tick time distribution and \newmetric during TNT workload on various AWS node sizes. Whiskers extend to $\pm$ 1.5 $\times$ IQR, bounded by the minimum and maximum values. Black diamond indicates arithmetic mean.}
    \label{fig:node_sizes}
    \vcutL{}
\end{figure}

Recommended hardware configurations in cloud environments result in unacceptable levels of performance variability, which degrades player experience.
By using more powerful cloud hardware, performance variability can be limited to acceptable levels.
Figure~\ref{fig:node_sizes} shows this result, showing both the mean tick duration and \newmetric for varying VM sizes in AWS.
We use the notation \emph{2XL}, \emph{XL}, \emph{L} to denote AWS VM sizes~\emph{t3.large}, \emph{t3.xlarge}, and \emph{t3.2xlarge} respectively.

Companies that specialize in cloud hosting of \mves commonly list recommended hardware configurations, with the most frequent recommendation being 2\,vCPUs and 4\,GB memory.
An overview of these recommendations is available in Table~\ref{tab:hardware_recs}.
These recommended values are significantly lower than those listed on the community-driven Minecraft wiki, which recommends a dedicated full CPU (e.g., Intel i5 or i7, or AMD Ryzen 5 or 7) and 6\,GB memory~\cite{serverrequirements}. This indicates that players experienced performance problems with the recommended hardware configuration.


Figure~\ref{fig:node_sizes} shows that using the recommended hardware configuration as listed by cloud-hosting companies,
which corresponds to the \emph{L} node type, results in poor performance and significant performance variability.
On this node size, each \mve becomes significantly overloaded by \envworkloads and exhibits high performance variability.

The larger node types \emph{XL} and \emph{2XL} have 4 and 8 vCPUs respectively~\cite{amazont3}.
While \emph{XL} provides better performance and less performance variability than \emph{L}, it remains insufficient to keep the mean tick time below 50\,ms.
The \emph{2XL} node type is required to provide sufficiently low mean tick duration.
However, this node type still shows significant performance variability for Minecraft (green cross) and Forge (blue square),
which means these games can still become overloaded temporarily, as shown in \Cref{sec:experiments:envworkloads-perf-variability}.



Interestingly, we observe that the benefit of more powerful hardware varies per \mve.
Specifically, while PaperMC's (red circle) performance instability (i.e., \newmetric) increases significantly when decreasing hardware resources, from 0.025 to 0.08 in the top-right sub-plot, it is the only game whose mean and 75th percentile tick duration stays well below the 50\,ms threshold.
Further analysis shows that, while PaperMC becomes overloaded and its tick duration exceeds 50\,ms, the number of extreme outliers is low, preventing this performance problem from becoming visible in the mean tick duration.

\section{Actionable Insights and Limitations}
\label{sec:actionable_insights}

The main findings in~\Cref{sec:experiments} lead to \textit{actionable insights}:


\begin{myitemize}[label=\textbf{I\arabic*}]
    \item\label{ai:reportvariability} Game developers and hardware producers should report performance variability when evaluating the performance of online games, using measures of variance such as \newmetricfull~(see \Cref{sec:metrics}) and the distribution of game response time and \fps. Games must provide consistently good performance to their users. Our experiments show that \mves can be overloaded and become unplayable, even when mean and median performance values indicate good performance~(\ref{mf:variability-player-experience}, Figure~\ref{fig:aws_response_time}).

    \item\label{ai:envworkloads} Game developers and hardware producers should include environment based workloads in their benchmarks for \mves. It is not sufficient to evaluate the performance of \mves using only large numbers of players (i.e., player-based workloads).
    \Envworkloads cause significant performance variability in \mves and make them unplayable~(\ref{mf:envworkloads-perf-variability}, Figure~\ref{fig:paw_jitter_scatter}), and must therefore be included in \mve benchmarks.


    \item\label{ai:bestcloud} Players should choose their cloud environment depending on their \mve, and should consider self-hosting their game. Our results indicate that choice of best cloud provider depends on the \mve. Minecraft and Forge obtain lower performance variability on AWS,
    while PaperMC obtains lower performance variability on Azure~(\ref{mf:cloud-perf-variability}, Figure~\ref{fig:50_iter_jitter}). Moreover, self-hosting remains a valuable alternative, resulting in significantly lower performance variability overall.

    \item\label{ai:hardwarereqs} \mve service providers should increase their hardware recommendations.  Prior work has shown that when asked to estimate in advance the hardware requirements of a given program, users either pick a provided default configuration, or overestimate to an extreme value to avoid performance issues~\cite{userestimatesinaccurate,userestimatessp2,userestimateshpc}. We find that recommended hardware configurations for hosting \mves on cloud environments are insufficient~(\ref{mf:default-hardware-insufficient}) and conclude that users who employ the first strategy will experience decreased quality of service which may cause them to switch to a competing commercial cloud provider. 

    To prevent this, commercial cloud providers should update hardware recommendations in line with our findings in Figure~\ref{fig:node_sizes}: a 2-core size is insufficient, and a node with 8~cores is necessary for smooth operation. The 4-core size provides a balance of cost and performance. Beyond these recommendations, commercial cloud providers should use our benchmark to determine adequate hardware allocations capable of fulfilling the service requirements of \mves under realistic workloads, and further adapt resource scheduling to be aware of the performance patterns of \mves. 

    Similarly, users who seek to avoid adverse performance variability when operating \mve cloud environments
    should choose node sizes comparable to the 8~core t3.2xlarge node, or use our benchmark to compare both various cloud
    providers and the specific \mve implementations.

    \item\label{ai:engineering} Game developers should engineer \mves to reduce impact of \envworkloads.
    Engineering for this goal can reduce the impact (i.e., performance variability) of environment-based workloads by 60\% on the same hardware, as shown by PaperMC operating the Farm workload on AWS~(Figure~\ref{fig:paw_jitter_scatter}).
    The goal of the PaperMC project is to implement a high performance \mve, including efficient processing of \envworkloads.
    We provide an analysis of PaperMC in Appendix~\ref{sec:appendix_a}.
    Developers creating \mves should consider and mitigate the impact of \envworkloads, and use our benchmark to measure, analyze and subsequently reduce the performance variability caused by such workloads. 

\end{myitemize}







\begin{addtext}
    Here we discuss limitations of our work 
    related to the \newmetricfull metric and the workloads.
    \newmetric cannot be used as a singular performance metric, but rather is designed as an additional axis by which to quantitatively appraise the performance of a game server, by capturing behavior that other metrics cannot. Because \newmetric is a measure of variability over an entire sampling duration, it does not capture extremely poor but stable performance, or the occurrence of singular relatively small outliers. Thus, other measures, such as percentiles, are necessary to observe the magnitude of tick durations and detect lone outliers. It is not currently understood how \newmetric directly relates to player-perceived quality of experience and quality of experience, and should be explored in future work, for instance through player studies.

    The workloads included in \toolname are intended to cover a wide range of realistic \envworkloads, from common to extreme cases. However, there is no publicly available analysis of \envworkload prevalence. Thus, we select \envworkloads
    using proxy metrics such as total views and downloads in online \mve communities,
    which may not be representative of all \mve players. Additionally, our experiments measuring the impact of \envworkloads utilized a purposefully minimal player-based workload component.
    Finally, because our experiments focused on \envworkloads, our
    player-based workloads (``Players") uses random avatar movement.
    Although real player behavior is likely more complex, no player-behavior models exists for \mves.
\end{addtext}
\section{Related Work}
\label{sec:relwork}
\label{sec:related-work}

We summarize in this section a developing overview of related work.
Overall, this study is the first to evaluate performance variability in \mves.
This is challenging because there is neither a generally accepted set of relevant workloads for \mves,
nor a standardized metric to quantify performance variability in computer systems.

Closest to our work, Yardstick is an \mve benchmark used to show the limited scalability of \mves~\cite{yardstick}.
The authors use Yardstick to evaluate the scalability and network characteristics of several \mve services.
However, Yardstick does not quantify performance variability, resulting in optimistic results.
Moreover, the authors do not evaluate \mve performance under \envworkloads or in the cloud. 

The MineRL competition~\cite{DBLP:journals/corr/abs-1907-13440} provides a dataset of Minecraft player
recordings. This dataset provides demonstrations to train artificial intelligence systems to complete a challenging
in-game task. In contrast,
the workloads used in this work focus on commonly observed patterns in the \mve community.

There exist several systems that aim to improve the scalability of \mves.
Manycraft~\cite{DBLP:conf/netgames/DiaconuKV13} increases the maximum number of players in a Minecraft instance by using Kiwano.
Kiwano~\cite{DBLP:conf/ieeehpcs/DiaconuK13} allows horizontal scaling of virtual environments through Voronoi partitioning, but requires a \emph{static environment}, which disables the \mve's modifiable world and is incompatible with \envworkloads.
Similar in many ways to Manycraft, Koekepan~\cite{DBLP:conf/netgames/EngelbrechtS13} uses zone-partitioning and scales horizontally.
Dyconits~\cite{DBLP:conf/icdcs/DonkervlietCI21} are a distributed architecture that scales \mves vertically, through the use of dynamically managed consistency-units. 
None of these approaches considers explicitly performance variability.


Iosup et al.~\cite{DBLP:conf/ccgrid/IosupYE11} find that commercial cloud environments exhibit significant yearly and daily performance variability patterns.
The authors show that performance variability varies per cloud operator, and use simulation experiments to show that this can affect the performance of applications, including a social online game.
In contrast, our benchmark uses real-world experiments to evaluate the effect of performance variability on \mves, which are real-time online games.

\section{Conclusion}
\label{sec:conclusion}

Online gaming is a popular and lucrative part of the entertainment industry, but raises important performance challenges.
In this work, we posit performance variability is an important cause for the lack of scalability
in \mves.


We make a four-fold contribution to better understand the behavior of \mves.
\textit{First}, we propose a novel workload model for these systems, which identifies important sources of performance variability not considered elsewhere. 
\textit{Second}, we design and implement \toolname, the first benchmark to evaluate performance variability in \mves.
\toolname uses realistic workload types; novel, it considers \envworkloads, and can evaluate \mves running both in self-hosted and cloud environments such as Amazon~AWS and Microsoft~Azure.
\textit{Third}, we use \toolname to perform real-world experiments and analyze the results.
We find that performance variability negatively affects players in \mves, that both \envworkloads and cloud environments can cause significant performance variability.
This leads us to formulate four actionable insights.
\textit{Fourth}, we release FAIR and FOSS artifacts that enable reproducibility for this work. 

In future work, we aim to conduct user studies to directly link our \newmetricfull~(\newmetric) values to player-perceived quality of experience.
To encourage community adoption, we aim to create a public score-board where operators of \mve{}-as-a-service can publish benchmark scores.




\bibliographystyle{ACM-Reference-Format}
\balance
\bibliography{references_beautified.bib}

\appendix
\section{The PaperMC Project}\label{sec:appendix_a}
Throughout all experiments it is seen that PaperMC performs better than either Minecraft or Forge. This merits an explanation of the differences between PaperMC and the other servers, such to drive future research into how to minimize the impact of \envworkloads through performance engineering. 

Much like Forge, the PaperMC project~\cite{papermc} is an open source set of community \textit{'patches'} applied on top of another server type (Spigot~\cite{spigot}) which in turn were modifications made to the Mojang-provided Minecraft server. Unlike Forge, PaperMC is not made in order to facilitate mods but instead to deal with a plethora of community grievances about server experience. As such, PaperMC at its core is a set of code patches and a repackaging of the server. The set of patches is outlined in their Github page~\cite{papermcgit}, with the most important being asynchronous threads and reworked thread priorities, limited per-thread cache duplication, a built-from scratch
scheduler, custom player defined logic algorithms, and importantly, an entirely new entity logic handler (as is found in~\ref{mf:entities-expensive}, entities are computationally expensive).

For how these changes improve performance, we compare the CPU flamegraph of the Vanilla and PaperMC server during the TNT workload, collected using perf-map-agent~\cite{perf-map-agent}. The flamegraph of Forge is omitted as it is (in this case) identical to Vanilla's. Flamegraphs are read as follows: the y-axis shows stack depth, the width of each box shows the total (not necessarily consecutive) time a given thread was on CPU. The ordering of boxes on the x-axis is arbitrary, in this case they are in alphabetical order. 

\begin{figure*}[t]
    \includegraphics[width=\linewidth]{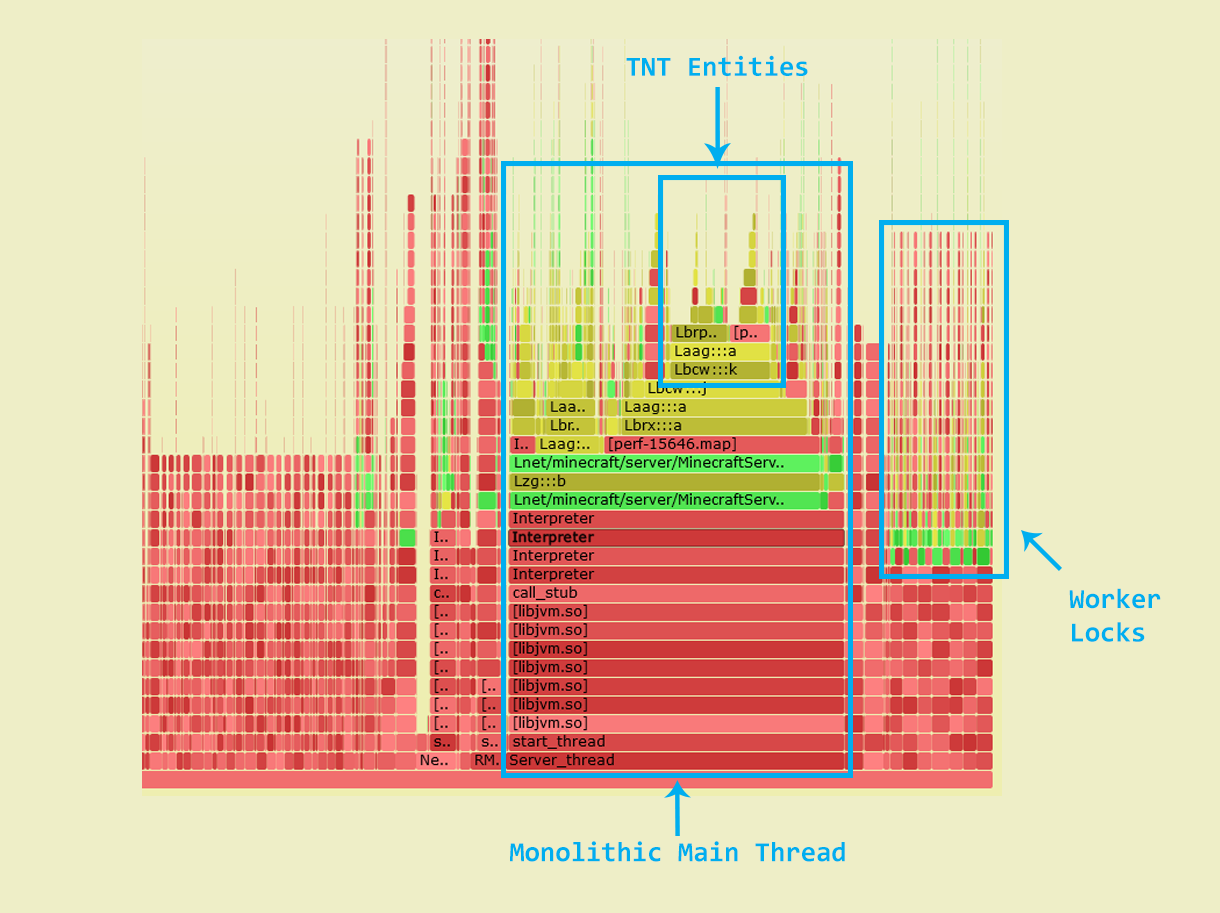}
    \caption{Minecraft flamegraph during TNT workload.}
    \label{fig:fg_vanilla}
\end{figure*}
\begin{figure*}[t]
    \includegraphics[width=\linewidth]{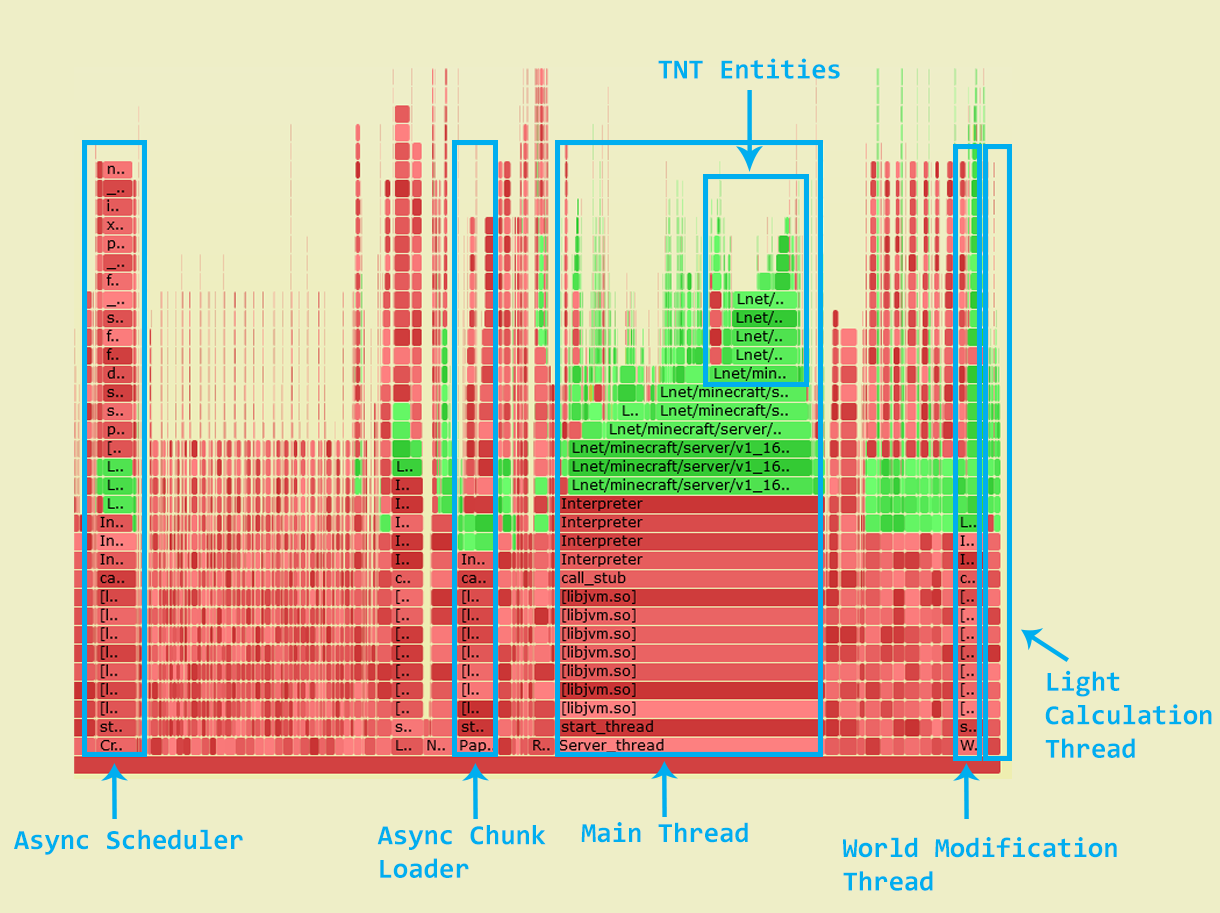}
    \caption{PaperMC flamegraph during TNT workload.}
    \label{fig:fg_papermc}
\end{figure*}

Figures~\ref{fig:fg_vanilla} and \ref{fig:fg_papermc} show that PaperMC hoists a number of environment operations out of the main thread into dedicated threads handled by a separate asynchronous scheduler thread. Thus the amount of time spent on worker locks is decreased, minimizing total resource contention. It can be extrapolated that these changes contribute strongly to PaperMC remaining less overloaded compared to Vanilla and Forge during the TNT workload.

\end{document}